\def\twocolumnversion{1}

\ifdefined \twocolumnversion
\documentclass[aps,pre,twocolumn,groupedaddress]{revtex4-1}
\else
\documentclass[aps,pre,groupedaddress]{revtex4-1}
\fi

\usepackage{amsmath} 
\usepackage{amssymb}  
\usepackage{amsthm}
\usepackage{color}
\usepackage{subfig}
\usepackage{hhline}

\newcommand{\dN}{\,dN}
\usepackage{accents}
\newcommand{\ubar}[1]{\underaccent{\bar}{#1}}
\DeclareMathOperator*{\col}{col}
\renewcommand{\Pr}{P}
\DeclareSymbolFont{bbold}{U}{bbold}{m}{n}
\DeclareSymbolFontAlphabet{\mathbbold}{bbold}
\newcommand{\onev}{\mathbbold{1}}

\usepackage{calrsfs}
\DeclareMathAlphabet{\pazocal}{OMS}{zplm}{m}{n}
\renewcommand{\mathcal}[1]{\pazocal{#1}}

\DeclareMathOperator*{\minimize}{minimize}
\DeclareMathOperator*{\subjectto}{subject\ to}

\newtheorem{definition}{Definition}

\newtheorem{problem}[definition]{Problem}
\usepackage{mathtools}

\usepackage[english]{babel}
\usepackage{blindtext}

\newcommand{\blue}[1]{#1}

\ifdefined \twocolumnversion

\else
\usepackage{changepage}
\fi

\begin{document}


\title{Epidemic Processes over Adaptive State-Dependent Networks}


\author{Masaki Ogura}
\email[]{ogura@seas.upenn.edu}
\author{Victor M.~Preciado}
\email[]{preciado@seas.upenn.edu}
\affiliation{University of Pennsylvania. 3330 Walnut Street, Philadelphia, PA 19104. USA.}

\date{\today}

\begin{abstract}
In this paper, we study the dynamics of epidemic processes taking place in
adaptive networks of arbitrary topology. We focus our study on the adaptive
susceptible-infected-susceptible (ASIS) model, where healthy individuals are
allowed to temporarily cut edges connecting them to infected nodes in order to
prevent the spread of the infection. In this paper, we derive a closed-form
expression for a lower bound on the epidemic threshold of the ASIS model
in arbitrary networks with heterogeneous node and edge dynamics. For networks
with homogeneous node and edge dynamics, we show that the resulting \blue{lower
bound} is proportional to the epidemic threshold of the standard SIS model over
static networks, with a proportionality constant that depends on the adaptation rates. Furthermore, based on our results, we propose an efficient algorithm to optimally tune the adaptation rates in order to eradicate epidemic
outbreaks in arbitrary networks. We confirm the \blue{tightness of the proposed
lower bounds} with several numerical simulations and compare our optimal
adaptation rates with popular centrality measures.
\end{abstract}


\maketitle


\section{Introduction}
 
The analysis of dynamic processes taking place in complex networks is a major
research area with a wide range of applications in social, biological, and
technological systems \cite{Watts1998, Newman2006, BBV08}. The spread of
information in online social networks, the evolution of an epidemic outbreak in
human contact networks, and the dynamics of cascading failures in the electrical
grid are relevant examples of these processes. While major advances have been
made in this field, most modeling and analysis techniques are specifically
tailored to study dynamic processes taking place in static networks.  However,
empirical observations in social \cite{Eagle2006, Cattuto2010, Santoro2011},
biological \cite{Przytycka2010,Han2004,Taylor2009}, and financial networks
\cite{Sabatelli2002} illustrate how real-world networks are constantly evolving
over time \cite{Holme2012}. Unfortunately, the effects of temporal structural
variations in the dynamics of networked systems remain poorly understood.

In the context of temporal networks, we are specially interested in the
interplay between the dynamics on networks (i.e., the dynamics of processes
taking place in the network) and the dynamics of networks (i.e., the temporal
evolution of the network structure). Although the dynamics on and of networks
are usually studied separately, there are many cases in which the evolution of
the network structure is heavily influenced by the dynamics of processes taking
place in the network. One of such cases is found in the context of epidemiology,
since healthy individuals tend to avoid contact with infected individuals in
order to protect themselves against the disease---a phenomenon called
\emph{social distancing}~\cite{Bell2006,Funk2010}. As a consequence of social
distancing, the structure of the network adapts to the dynamics of the epidemics
taking place in the network. Similar adaptation mechanisms have been studied in
the context of power networks~\cite{Scire2005}, biological and neural
networks~\cite{Hopfield1983,Schaper2003} and on-line social
networks~\cite{Antoniades2013}.

Despite the relevance of network adaptation mechanisms, their effects on the
network dynamics are not well understood. In this research direction, we find
the seminal work by Gross et al.~in \cite{Gross2006}, where a simple adaptive
rewiring mechanism was proposed in the context of epidemic models. In this
model, a susceptible node can cut edges connecting him to infected neighbors and
form new links to \emph{any} randomly selected susceptible nodes---without
structural constraint in the formation of new links. Despite its simplicity,
this adaptation mechanism induces a complex bifurcation diagram including
healthy, oscillatory, bistable, and endemic epidemic states \cite{Gross2006}.
Several extensions of this work can be found in the
literature~\cite{Gross2008,Zanette2008a,Marceau2010,Lagorio2011,Rogers2012a,Juher2013,Tunc2014}, where the authors assume homogeneous infection and recovery rates in the network. Another model that is specially relevant to our work is the adaptive susceptible-infected-susceptible (ASIS) model proposed in~\cite{Guo2013}. In this model, edges in a given contact network can be temporarily removed in order to prevent the spread of the epidemic. An interesting feature of the ASIS model is that, in contrast with Gross' model, it is able to account for arbitrary contact patterns, since links are constrained to be part of a given contact graph. Despite its modeling flexibility, analytical results for the ASIS model~\cite{Guo2013,Trajanovski2015} are based on the assumption of homogeneous contact patterns (i.e., the contact graph is complete), as well as homogeneous node and edge dynamics (i.e., nodes present the same infection and recovery rates, and edges share the same adaptation rates).

As a consequence of the lack of tools to analyze network adaptation mechanisms,
there is also an absence of effective methodologies for actively utilizing
adaptation mechanisms for containing spreading processes. Although we find in
the literature a few attempts in this direction, most of them rely on extensive
numerical simulations~\cite{Bu2013}, on assuming a homogeneous  contact patterns
\cite{Maharaj2012}, or a homogeneous node and edge dynamics~\cite{Valdez2012}.
In contrast, while controlling epidemic processes over static networks, we find
a plethora of tools based on game theory~\cite{Trajanovski2015a,Aspnes2006} or
convex optimization~\cite{Preciado2014,Nowzari2015a}.

In this paper, we study adaptation mechanisms over arbitrary contact networks.
In particular, we derive an explicit expression for a \blue{lower bound} on the
epidemic threshold of the ASIS model for arbitrary networks, as well as
heterogeneous node and edge dynamics. In the case of homogeneous node and edge
dynamics, we show that the \blue{lower bound} is proportional to the epidemic
threshold of the standard SIS model over a static network
\cite{VanMieghem2009a}. Furthermore, based on our results, we propose an
efficient algorithm for optimally tuning the adaptation rates of an arbitrary
network in order to eradicate an epidemic outbreak in the ASIS model. We confirm
the \blue{tightness of the proposed lower bonds} with several numerical
simulations and compare our optimal adaptation rates with popular centrality
measures.

\section{Heterogeneous ASIS Model} 

In this section, we describe the adaptive susceptible-infected-susceptible
(ASIS) model over \emph{arbitrary} networks with \emph{heterogeneous} node and
edge dynamics (heterogeneous ASIS model for short). We start our exposition by
considering a spreading process over a time-varying contact graph~$\mathcal
G(t)=(\mathcal V,\mathcal E(t))$, where $\mathcal V=\{1,\ldots,n\}$ is the set
of nodes and $\mathcal E(t)$ is the time-varying set of edges. For any $t\geq
0$, $A(t) = [a_{ij}(t)]_{i,j}$ corresponds to the adjacency matrix of $\mathcal
G(t)$, and the neighborhood of node~$i$ at time $t$ is defined as $\mathcal
N_i(t) =\{j: \{i,j\}\in\mathcal E(t)\}$. In the standard SIS epidemic model, the
state of node $i$ at time $t$ is described by a Bernoulli random variable
$x_i(t)\in\{0, 1\}$, where node~$i$ is said to be \emph{susceptible} if $x_i(t)
= 0$, and \emph{infected} if $x_i(t) = 1$.  When the contact graph evolves over
time, the evolution of $x_i$ is described by a Markov process with the following
transition probabilities:
\begin{align}
&\begin{multlined}[c][.8\linewidth]
\Pr(x_i(t+h) = 1 \mid x_i(t) = 0) =\beta_i\ \sum_{\mathclap{k \in \mathcal N_i(t)}}\ x_k(t)\,h + o(h),
\end{multlined} \nonumber 
\\
&\Pr(x_i(t+h) = 0 \mid x_i(t) = 1) = \delta_i h + o(h), 
\label{eq:recovery}
\end{align}
where the parameters $\beta_i > 0$ and $\delta_i > 0$ are called the \emph{infection} and \emph{recovery} rates of node~$i$.

\begin{figure}[tb]
\centering
\includegraphics[width=.9\linewidth]{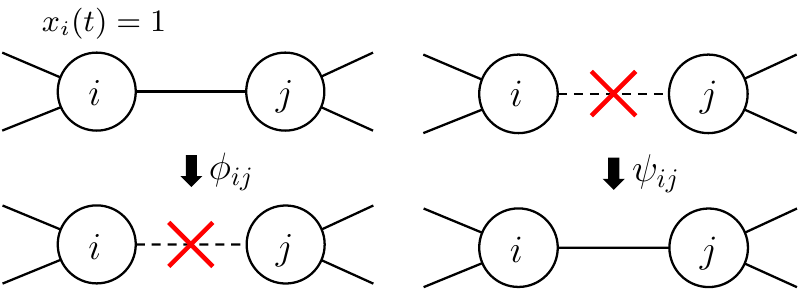}
\caption{Adaptation mechanisms in the ASIS model.}
\label{fig:adaptive}
\end{figure}

In the heterogeneous ASIS model, the epidemics takes place over a time-varying
network that we model as a continuous-time stochastic graph process~$\mathcal G
= \{\mathcal G(t) \}_{t\geq 0}$, described below. Let $\mathcal G(0)=(\mathcal
V,\mathcal E(0))$ be an initial connected contact graph with adjacency
matrix~$A(0) = [a_{ij}(0)]_{i,j}$. We assume that $\mathcal G(0)$ is strongly
connected. Edges in the initial graph $\mathcal G(0)$ appear and disappear over
time according to the following Markov processes:
\begin{align}
&\begin{multlined}[c][.9\linewidth]
\Pr(a_{ij}(t+h) = 0 \mid a_{ij}(t) = 1)=\vspace{.1cm}
\\
\phi_{ij}x_i(t) h+ \phi_{ji}x_j(t) h +  o(h),
\end{multlined}\label{eq:cut}
\\
&\Pr(a_{ij}(t+h) = 1 \mid a_{ij}(t) = 0) = a_{ij}(0)\psi_{ij} h +  o(h),
\label{eq:rewire}
\end{align}
where the parameters $\phi_{ij} > 0$ and $\psi_{ij} = \psi_{ji}> 0$ are called
the \emph{cutting} and \emph{reconnecting} rates. Notice that the transition
rate in \eqref{eq:cut} depends on $x_i$ and $x_j$, inducing an adaptation
mechanism of the network structure to the state of the epidemics. The transition
probability in~\eqref{eq:cut} can be interpreted as a protection mechanism in
which edge $\{i,j\}$ is stochastically removed from the network if either node
$i$ or $j$ is infected. More specifically, because of the first summand
(respectively, the second summand) in \eqref{eq:cut}, whenever node $i$ 
(respectively, node $j$) is infected, edge~$\{i, j\}$ is removed from the
network according to a Poisson process with rate~$\phi_{ij}$ (respectively,
rate~$\phi_{ji}$). On the other hand, the transition probability in
\eqref{eq:rewire} describes a mechanism for which a `cut' edge $\{i, j\}$ is
`reconnected' into the network according to a Poisson process with rate
$\psi_{ij}$ (see Figure~\ref{fig:adaptive}). Notice that we include the term
$a_{ij}(0)$ in \eqref{eq:rewire} to guarantee that only edges present in the
initial contact graph $\mathcal G(0)$ can be added later on by the reconnecting
process. In other words, we constrain the set of edges in the adaptive network
to be a part of the arbitrary contact graph $\mathcal G(0)$.

\section{Epidemic Thresholds} 

In this section, we derive a \blue{lower bound} on the epidemic threshold for the heterogeneous ASIS
model. For $\gamma > 0$, let $N_\gamma$ denote a Poisson counter with
rate~$\gamma$ \cite{Feller1956vol1}. In what follows, we assume all Poisson
counters to be stochastically independent. Then, from the two equations in
\eqref{eq:recovery}, the evolution of the nodal states can be described by the
following set of stochastic differential equations
\begin{equation}\label{eq:dx_i}
dx_i = -x_i \dN_{\delta_i} + (1-x_i) \ \sum_{\mathclap{k\in\mathcal N_i(0)}}\  a_{ik}x_k \dN_{\beta_i},
\end{equation}
for all~$i\in \mathcal V$. Similarly, from \eqref{eq:cut} and \eqref{eq:rewire}, the
evolution of the edges can be described by the following set of stochastic differential equations:
\begin{equation}\label{eq:da_ij}
da_{ij} = -a_{ij} (x_i \dN_{\phi_{ij}} + x_j \dN_{\phi_{ji}} )+ (1-a_{ij}) \dN_{\psi_{ij}},
\end{equation}
for all $\{i, j\}\in \mathcal E(0)$.

By \eqref{eq:dx_i}, the expectation $E[x_i]$ obeys the differential equation
\begin{equation*}
\frac{d}{dt}E[x_i] 
= 
-\delta_i E[x_i] + \beta_i \ \sum_{\mathclap{k\in\mathcal N_i(0)}}\  E[(1-x_i)a_{ik}x_k]. 
\end{equation*}
Let $p_i(t) = E[x_i(t)]$ and $q_{ij}(t) = E[a_{ij}(t)x_i(t)]$. Then, it follows
that
\begin{equation}\label{eq:dpidt}
\frac{dp_i}{dt}
= 
-\delta_i p_i + \beta_i \ \sum_{\mathclap{k\in\mathcal N_i(0)}}\ q_{ki}
-
f_i,\text{ for }i=1,\ldots,n,
\end{equation}
where $$f_i(t) = \beta_i \sum_{k\in\mathcal N_i(0)}E[x_i(t)x_k(t)a_{ik}(t)]$$
contains positive higher-order terms. In what follows, we derive a set of
differential equations to describe the evolution of $q_{ij}$. From
\eqref{eq:dx_i} and \eqref{eq:da_ij}, we obtain the following equation using
It\^o rule for jump processes (see, e.g., \cite{Hanson2007})
\begin{align*}
d(a_{ij}x_i) = & -a_{ij} x_i \dN_{\phi_{ij}} - a_{ij}x_ix_j\dN_{\phi_{ji}} \vspace{.1cm}
\\
  & +(1-a_{ij})x_i \dN_{\psi_{ij}} - a_{ij}x_i \dN_{\delta_i} 
\\
  & +a_{ij}(1-x_i)\ \sum_{\mathclap{k\in\mathcal N_i(0)}}\  a_{ik}x_k \dN_{\beta_i}. 
\end{align*}
Therefore, 
\begin{align}\label{eq:qijdt}
\frac{dq_{ij}}{dt} = & -\phi_{ij} p_{ij} +  \psi_{ij} (p_i - q_{ij}) \nonumber
\\ & - \delta_i q_{ij} + \beta_i \ \sum_{\mathclap{k\in\mathcal N_i(0)}}\ q_{ki}
- g_{ij},
\end{align}
for all $\{i,j\}\in \mathcal E(0)$, where 
\begin{align*}
g_{ij}(t)  = & \phi_{ji} E[x_i(t)x_j(t)a_{ij}(t)]\\
& +\beta_i\ \sum_{\mathclap{k\in\mathcal N_i(0)}}\ E\bigl[ x_i(t)x_k(t)a_{ik}(t) \\
& + (1-a_{ij}(t))a_{ik}(t)x_k(t)\bigr],
\end{align*}
which contains positive higher-order terms. The differential equations~\eqref{eq:dpidt} and
\eqref{eq:qijdt} describe the joint evolution of the spreading process and the network structure.

For further analysis, it is convenient to express the differential
equations~\eqref{eq:dpidt} and \eqref{eq:qijdt} using vectors and matrices. For
this purpose, let us introduce the following notation. Let $I_{r}$ and
$\onev_{s}$ be, respectively, the $r\times r$ identity matrix and the
$s$-dimensional column vector of all ones. Given two matrices $M_1$ and $M_2$,
their Kronecker product~\cite{Horn1990} is denoted by $M_1 \otimes M_2$. Given a
sequence of matrices~$A_1,\dotsc,A_n$, their direct sum, denoted by
$\bigoplus_{i=1}^n A_i$, is defined as the block diagonal matrix having 
$A_1,\dotsc,A_n$ as its block diagonals \cite{Horn1990}. If $A_1,\dotsc,A_n$
have the same number of columns, then the matrix obtained by stacking
$A_1,\dotsc,A_n$ in vertical ($A_1$ on top) is denoted by $\col_{1\leq i\leq n}
A_i = \col(A_1, \dotsc, A_n)$. Based on this notation, we define the
vector-variable $p = \col_{1\leq i\leq n}p_i$, which contains the infection
probabilities of all the nodes in the graph. Similarly, let $q_i = \col_{j\in
\mathcal N_i(0)}q_{ij}$ and define the column vector $q = \col_{1\leq i\leq n}
q_i$. Define $T_i$ as the unique row-vector satisfying
\begin{equation}\label{eq:def:Ti}
T_i q = \sum_{{k \in \mathcal N_i(0)}} q_{ki}.
\end{equation}
Note that the length of the row vector $T_i$ and the column vector $q$ equals $2m$, where $m$ is the number of the edges in the initial graph $\mathcal G(0)$. 

Using this notation, we define the following matrices:
\begin{equation*}
\begin{gathered}
\begin{aligned}
B_1 = &
\begin{bmatrix}
\beta_1 T_1
\\
\vdots
\\
\beta_n T_n
\end{bmatrix},
&
B_2 = &
\begin{bmatrix}
\beta_1 \onev_{d_1} \otimes  T_1
\\
\vdots
\\
\beta_n \onev_{d_n} \otimes  T_n
\end{bmatrix}, 
\\
D_1 = & \bigoplus_{i=1}^n\delta_i,\ 
&
D_2 = & \bigoplus_{i=1}^n( \delta_i I_{d_i}), 
\end{aligned}
\end{gathered}
\end{equation*}
where  $d_i$  denotes the degree of node~$i$ in the initial graph~$\mathcal G(0)$. Furthermore, we also define the following matrices
\begin{equation*}
\begin{gathered}
\Phi = \bigoplus_{i=1}^n \bigoplus_{j \in \mathcal N_i(0)} \phi_{ij},
\\
\Psi_1 = \bigoplus_{i=1}^n \left(\col_{j\in{\mathcal N}_i(0)}\psi_{ij}\right),\ 
\Psi_2 = \bigoplus_{i=1}^n 
\;\;\bigoplus_{\mathclap{j\in\mathcal N_i(0)}}\;\psi_{ij}.
\end{gathered}
\end{equation*}

Stacking the set of $n$ differential equations in \eqref{eq:dpidt} into a single vector equation, and ignoring the negative higher-order term $-f_i$, we obtain the following entry-wise vector inequality for the probabilities of infection:
\begin{equation}\label{eq:dpdt<=...}
\frac{dp}{dt}\leq -D_1 p + B_1 q. 
\end{equation}
Also, stacking the set of differential equations in \eqref{eq:qijdt} with respect to $j\in
\mathcal N_i(0)$, and ignoring the negative term~$-g_{ij}$, we obtain the following entry-wise vector inequality: 
\begin{equation*}
\frac{dq_i}{dt} 
\leq
\col_{{j\in\mathcal N_i(0)}} (\psi_{ij}p_i) - 
(\phi_{ij} + \delta_i) q_i 
- \psi_j q_i + \beta_i( \onev_{d_i}\otimes T_i ) q, 
\end{equation*}
where $\psi_i = \bigoplus_{j\in\mathcal N_i(0)} \psi_{ij}$. We can further stack the above inequalities with respect to the index $i$ to obtain the following entry-wise vector inequality:
\begin{equation*}
\frac{dq}{dt} \leq \Psi_1 p + (B_2 -D_2-\Phi - \Psi_2)q.
\end{equation*}
Combining this inequality and \eqref{eq:dpdt<=...}, we obtain
\begin{equation}\label{eq:d[pq]/dt}
\frac{d}{dt}\begin{bmatrix}
p \\ q
\end{bmatrix} 
\leq
M 
\begin{bmatrix}
p \\ q
\end{bmatrix}, 
\end{equation}
where $M$ is an irreducible matrix (see Appendix~\ref{appx:pf:irreducibility} for the proof of irreducibility) defined as
\begin{equation}\label{eq:defM}
M  = 
\begin{bmatrix}
-D_1	&B_1
\\
\Psi_1	&B_2 - D_2 - \Phi - \Psi_2
\end{bmatrix}.
\end{equation}
Therefore, the evolution of the joint vector variable~$\col(p,q)$ is
upper-bounded by the linear dynamics given by the matrix $M$. Moreover, the
upper bound is tight around the origin, since both $f_i$ and $g_{ij}$ consist of
higher-order terms. From \eqref{eq:defM}, we conclude that the epidemics dies
out exponentially fast in the heterogeneous ASIS model if
\begin{equation}\label{eq:thr}
\lambda_{\max}(M) < 0, 
\end{equation}
where $\lambda_{\max}(M)$ is defined as the maximum among the real parts of the eigenvalues of $M$. Furthermore, since  $M$ is a Metzler matrix (i.e., has nonnegative off-diagonals) and irreducible, there is a real eigenvalue of $M$ equal to~$\lambda_{\max}(M)$ \cite{Horn1990}.

In the homogeneous case, where all the nodes share the same infection rate
$\beta > 0$ and recovery rate~$\delta>0$, and all the edges share the same
cutting rate~$\phi > 0$ and reconnecting rate~$\psi > 0$, the
condition~\eqref{eq:thr} reduces to the following inequality:
\begin{equation}\label{eq:threshold:homo}
\frac{\beta}{\delta} < \frac{1 + \omega}{\rho}, 
\end{equation}
where $\rho$ is the spectral radius of the initial graph $\mathcal G(0)$ and
$$\omega = \frac{\phi}{\delta+\psi},$$
which we call the \emph{effective cutting rate}. The proof of the extinction
condition \eqref{eq:threshold:homo} is given in
Appendix~\ref{appx:pf:thm:stbl:homo}. We remark that, in the special case when
the network does not adapt to the prevalence of infection, i.e., when $\phi=0$,
we have that $\omega = 0$ and, therefore, the condition
in~\eqref{eq:threshold:homo} is identical to the extinction
condition~$\beta/\delta<1/\rho$ corresponding to the homogeneous networked SIS
model over a static network~\cite{VanMieghem2009a}.

\blue{It is worth comparing the condition in \eqref{eq:threshold:homo} with the
epidemic threshold $\tau_c$ given in \cite{Guo2013} for the case in which $\mathcal G(0)$ is the complete graph:
\begin{equation}\label{eq:guo}
\tau_c = \frac{\omega_1 - 1}{n(h(\omega_1; \xi/\delta) - 2 + n^{-1})}, \quad \omega_1 = \frac{2\zeta}{\xi}, 
\end{equation}
where $\xi$ is the link-creating rate, $\zeta$ is the link-breaking rate, and
$h$ is a positive and ``slowly varying'' function depending on the metastable
long-time average of the number of infected nodes (for details, see
\cite{Guo2013}). We first notice that our \blue{lower bound} in
\eqref{eq:threshold:homo} can be checked directly from the parameters of the
model, namely, the adjacency matrix of the initial graph $\mathcal G(0)$ and the
relevant rates of the model. This is in contrast with the threshold
in~\eqref{eq:guo}, since it depends on the metastable average of the number of
infected nodes and, thus, can only be computed via numerical simulations. We
also remark that the lower bound on the epidemic threshold in
\eqref{eq:threshold:homo} and the epidemic threshold in \eqref{eq:guo} both
exhibit affine dependence on the effective link-breaking rates~$\omega$
and~$\omega_1$, respectively. Finally, we see that the recovery rate $\delta$
appears in different places in the two conditions, namely, inside the expression
of $\omega$ in \eqref{eq:threshold:homo} and inside the function~$h$ in
\eqref{eq:guo}. However, the consequences of this difference are not obvious,
since $h$ is defined via the metastable state and, therefore, does not allow an
analytical investigation.}

\begin{table*}[tb]
\newcommand{\figwidth}{.29\textwidth}
\newcommand{\gwidth}{.09\textwidth}
\newcommand{\gheight}{4cm}
\caption{$y^*$ versus $\beta$ and $\phi$, with $\delta = 1$. The dashed straight lines show the analytically derived \blue{lower bound} $(1+\omega)/\rho = \beta$ on the epidemic threshold. The contour plots are obtained by spline-interpolations of the discrete data obtained from the simulations.}
\label{table:new}
\begin{tabular}{cccc}
\hline	
& a) $\psi=1/2$ & {b) $\psi=1$} & c) $\psi=2$
\\
\hhline{====}
\parbox[c][\gheight]{\gwidth}{1) Erd\H{o}s-\\R\'enyi\\graph\\($p=0.1$)} &
\parbox[c]{\figwidth}{\includegraphics[width=\figwidth]{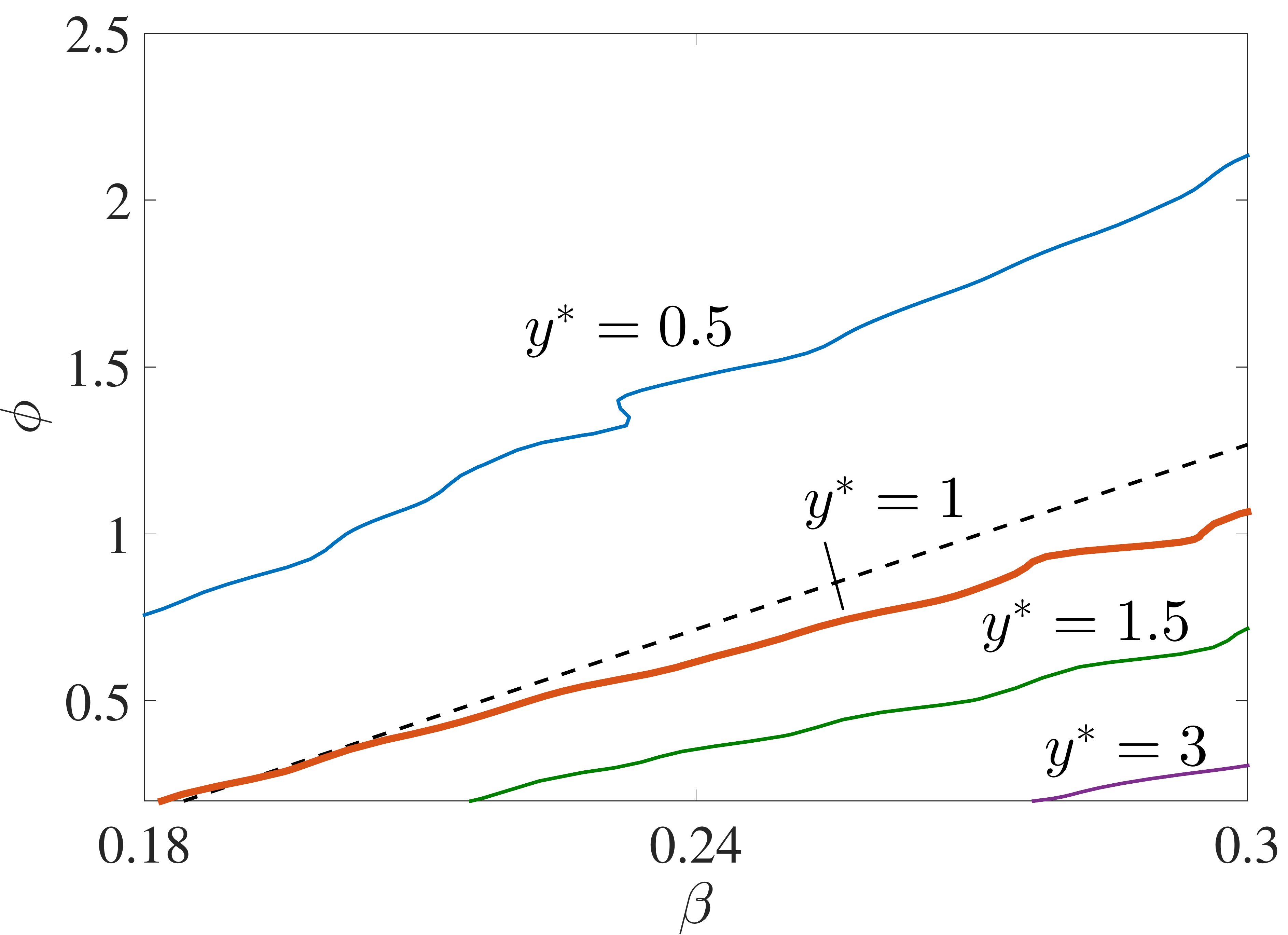}}&
\parbox[c]{\figwidth}{\includegraphics[width=\figwidth]{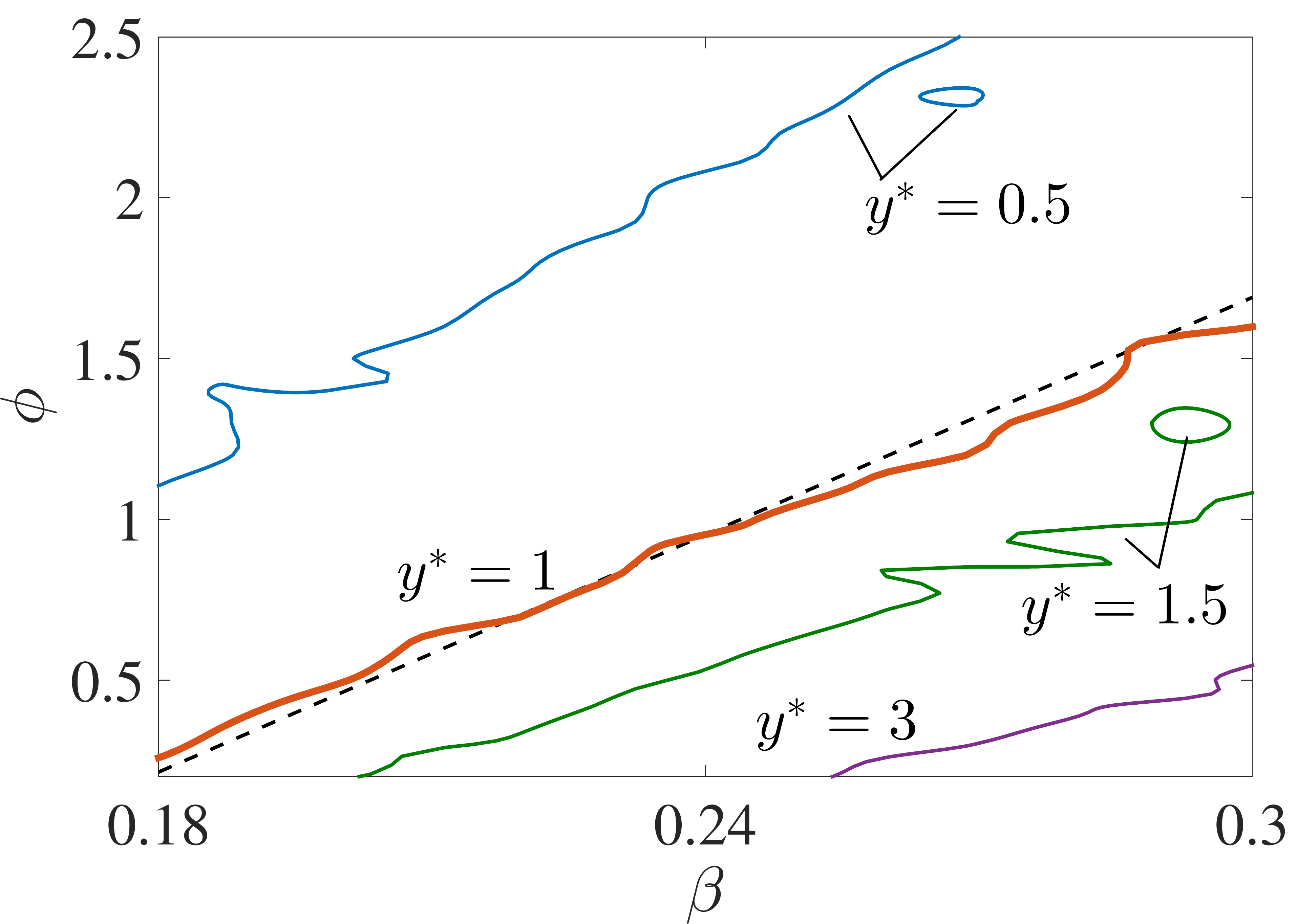}} &
\parbox[c]{\figwidth}{\includegraphics[width=\figwidth]{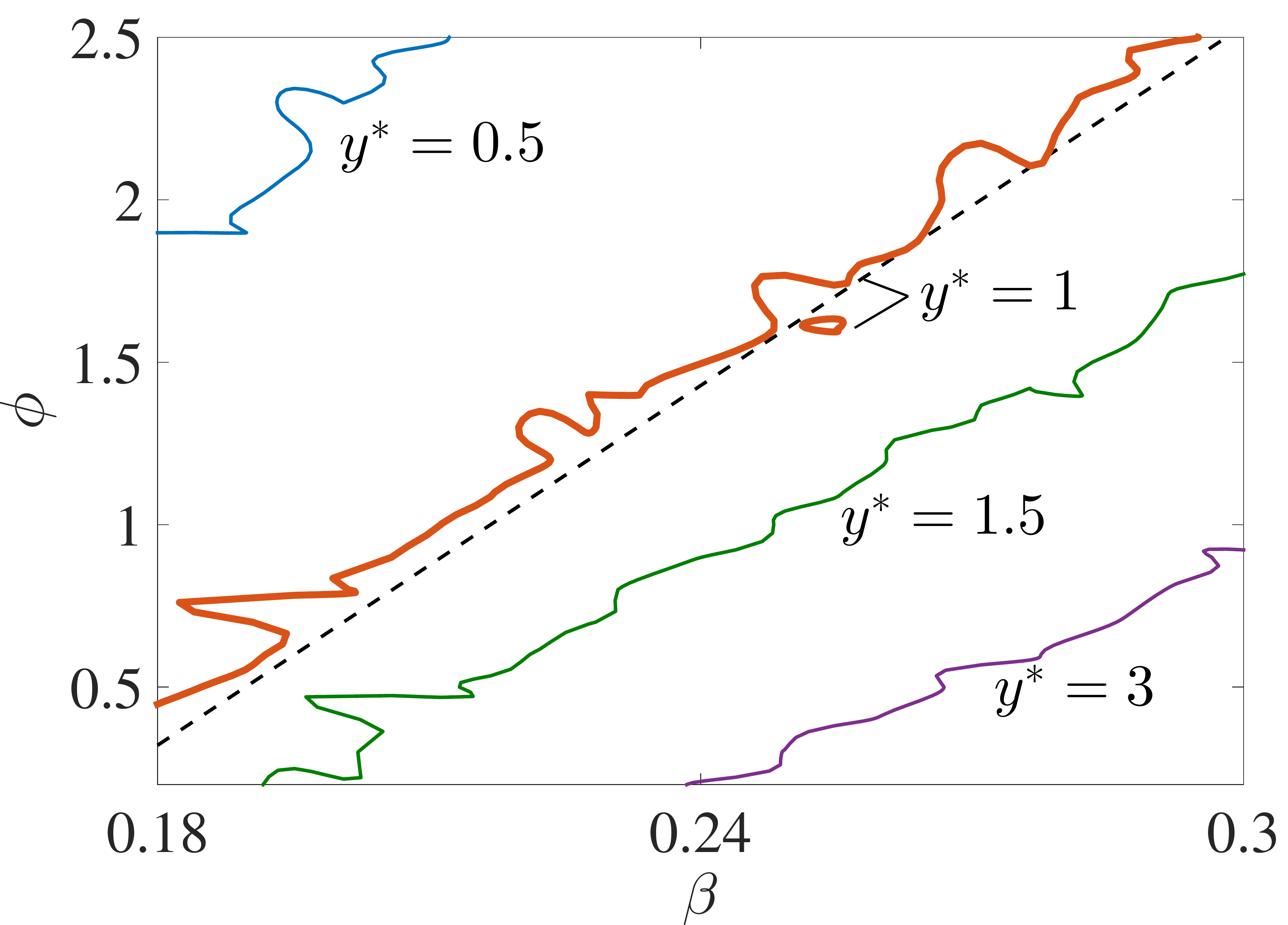}}
\\
\hline
\parbox[c][\gheight]{\gwidth}{2) Erd\H{o}s-\\R\'enyi\\graph\\($p=0.2$)} &
\parbox[c]{\figwidth}{\includegraphics[width=\figwidth]{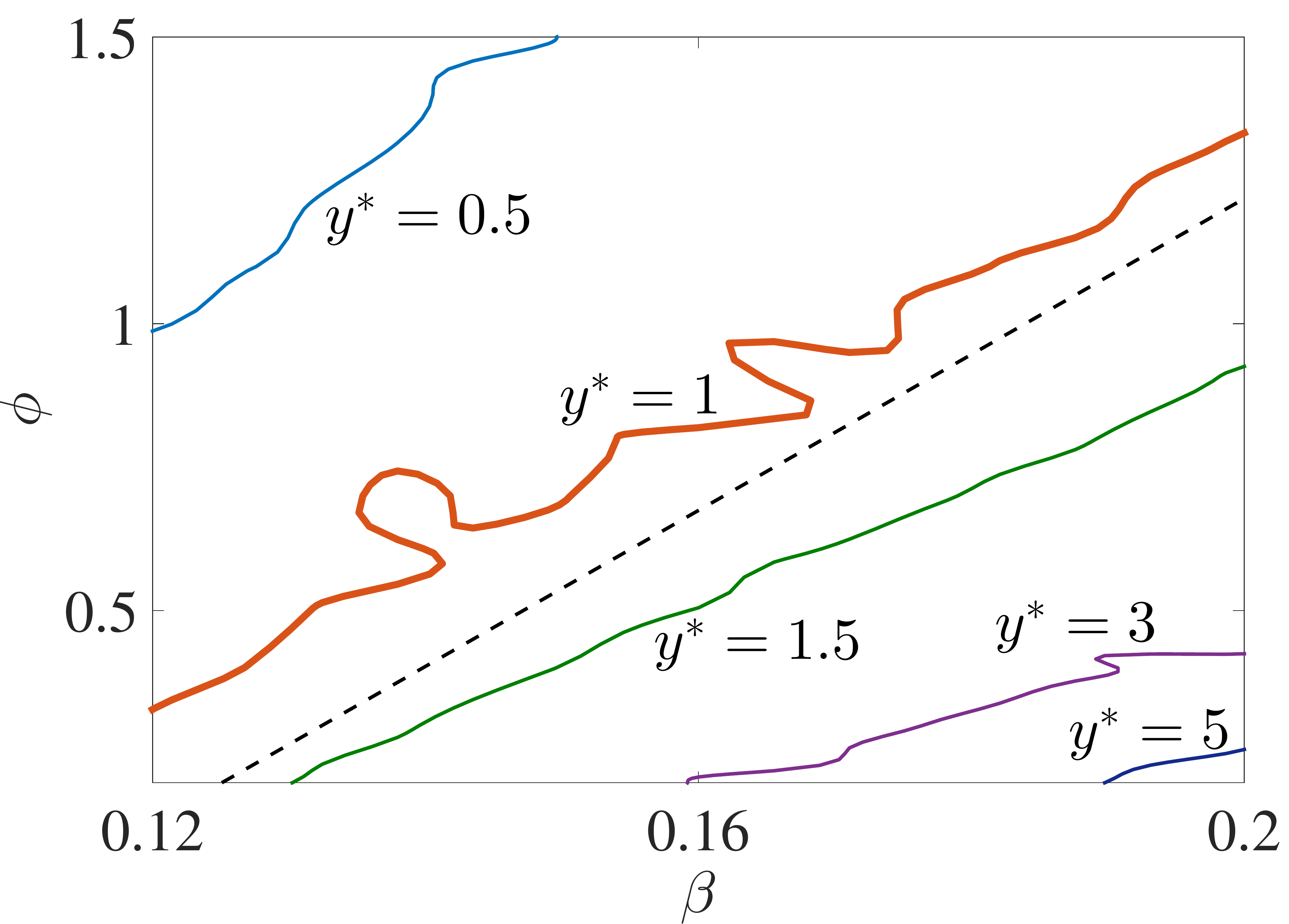}}&
\parbox[c]{\figwidth}{\includegraphics[width=\figwidth]{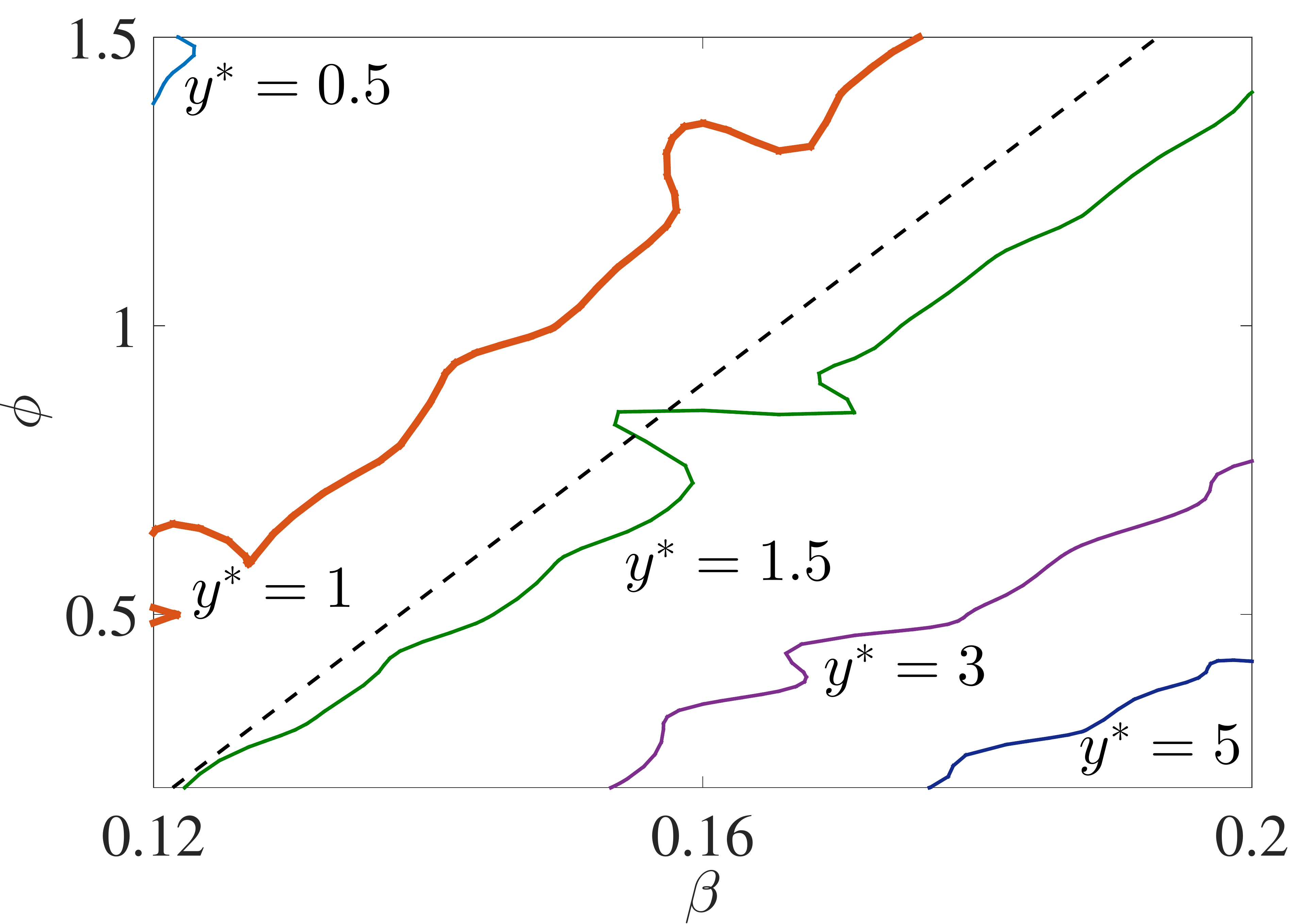}} &
\parbox[c]{\figwidth}{\includegraphics[width=\figwidth]{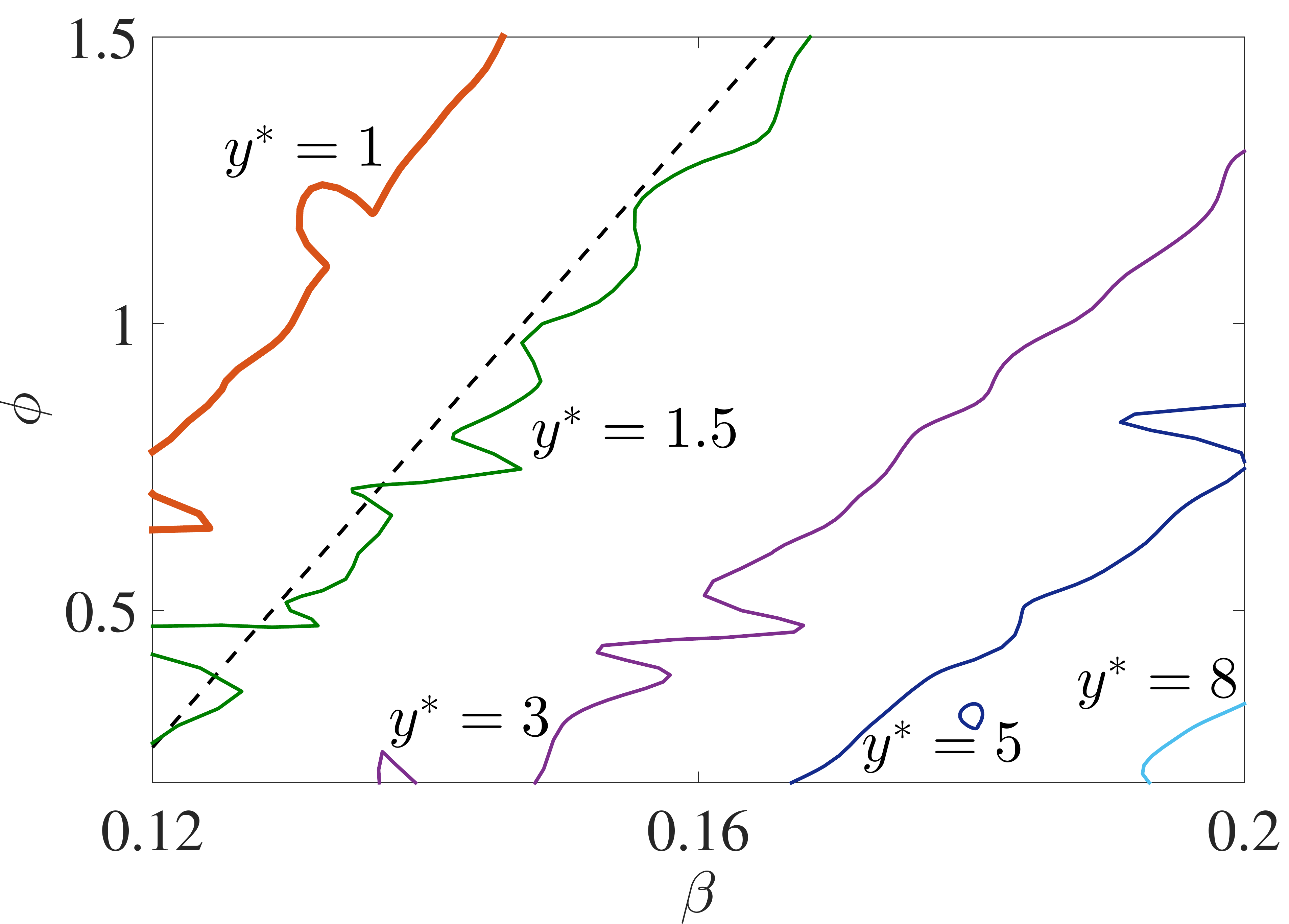}}
\\
\hline
\parbox[c][\gheight]{\gwidth}{3)~Barab\`asi-Albert\\model} &
\parbox[c]{\figwidth}{\includegraphics[width=\figwidth]{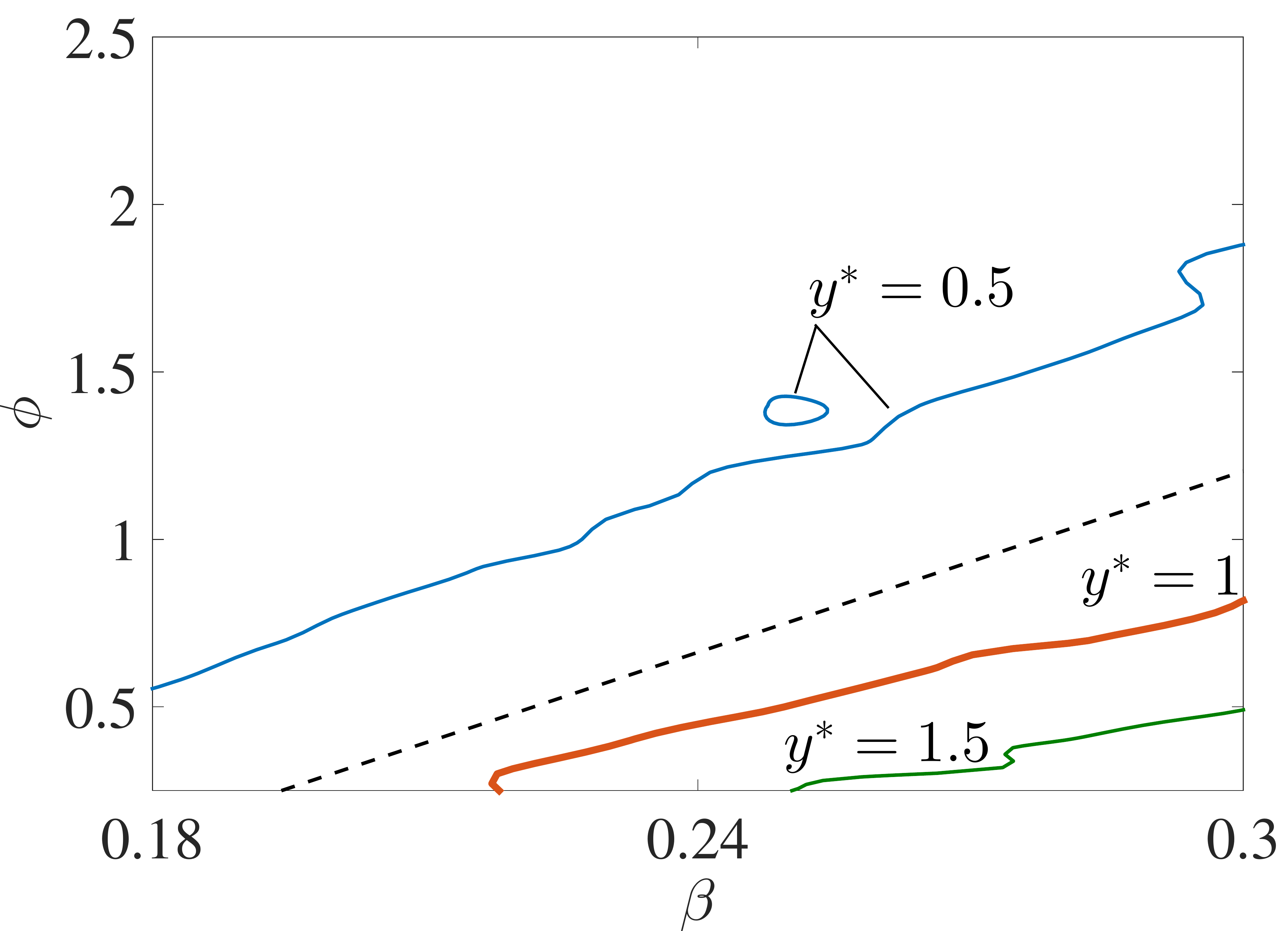}}&
\parbox[c]{\figwidth}{\includegraphics[width=\figwidth]{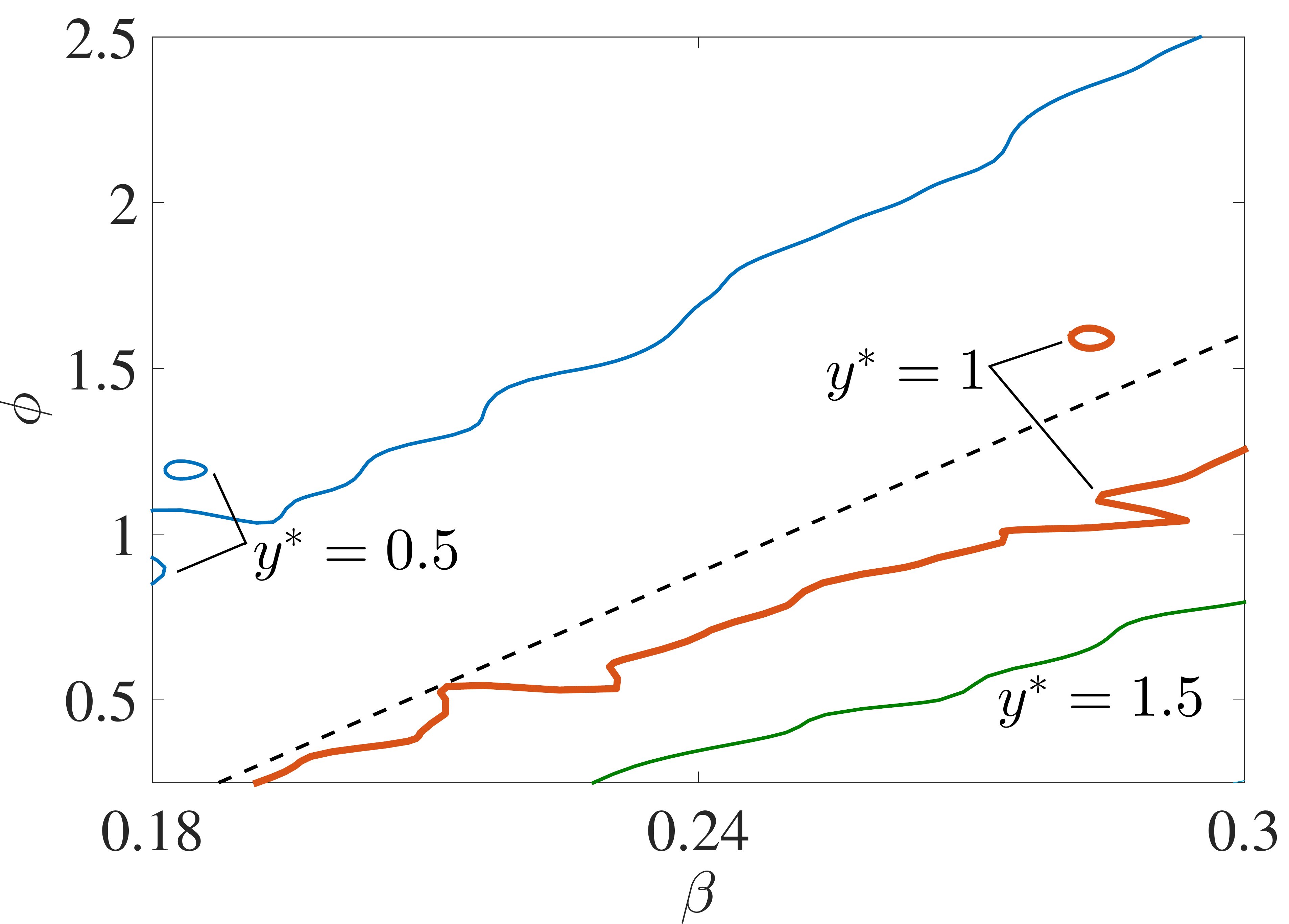}} &
\parbox[c]{\figwidth}{\includegraphics[width=\figwidth]{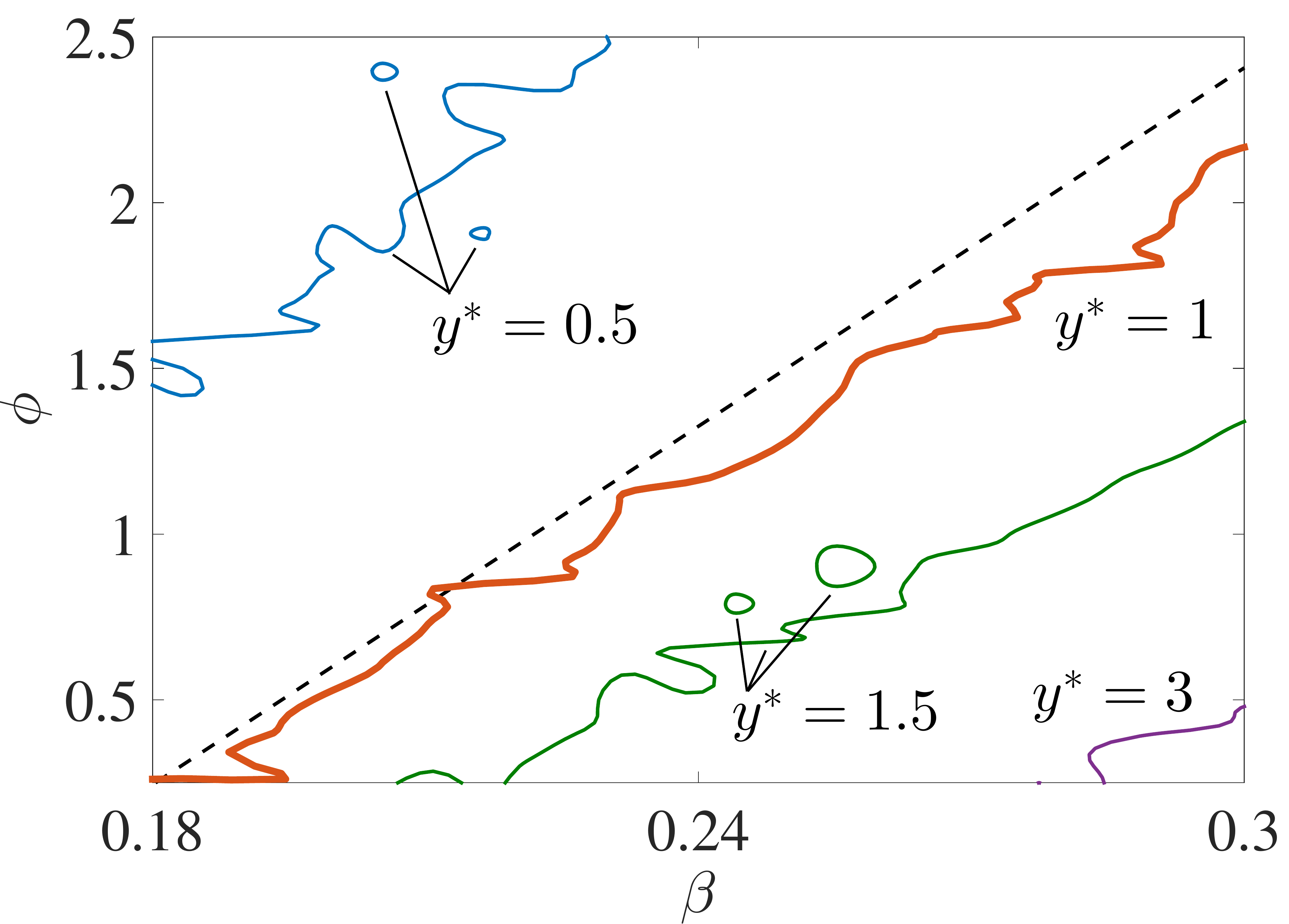}}
\\
\hline
\end{tabular}
\end{table*}

We now check the tightness of the \blue{lower bound}
in~\eqref{eq:threshold:homo} with numerical simulations. To find the metastable
number of infected nodes, we compute the long-time average of the number of
infected nodes, defined as
\begin{equation}
y(t) = \frac{1}{t}\int_0^t \sum_{i=1}^n x_i(\tau) \,d\tau \label{eq:y(t)}
\end{equation}
(for a sufficiently large $t$). In practice, the epidemics can die out during the simulation due to random fluctuations. To prevent this from happening, we use the procedure in~\cite{Cator2013a}, where randomly chosen nodes are immediately reinfected after the infection process dies (i.e., when all the variables $x_1(t)$, $\dotsc$, $x_n(t)$ become zero). To make sure that the process has reached the metastable state, we use the method in~\cite{Cator2013a}, where two independent simulations  are simultaneously run on the same network. One simulation starts with a 10\% of randomly chosen infected nodes, whereas the second simulation starts with all the nodes infected. For each simulation, we compute the long-time average of infected nodes using \eqref{eq:y(t)}, which we denote by $y_1(t)$ and $y_2(t)$, respectively. Similarly, we compute the long-time average of the number of edges present in the networks using the expression $z(t) = t^{-1}\int_0^t \sum_{i<j}a_{ij}(\tau)\,d\tau$, for each one of the two simulations, which we denote by $z_1(t)$ and $z_2(t)$, respectively. Following the procedure in \cite{Cator2013a}, the simulation is stopped when the following condition is satisfied $$\frac{|y_1(t)-y_2(t)|}{y_1(t)+y_2(t)}+\frac{|z_1(t)-z_2(t)|}{z_1(t)+z_2(t)}<10^{-4}.$$Once the simulation is stopped, the metastable number of infected nodes is determined by 
\begin{equation*}
y^* = y(t) - 1, 
\end{equation*}
where the subtraction of one compensates the effect of the re-infection procedure used in our simulations.

\begin{figure*}[tb]
\includegraphics[width=.31\textwidth]{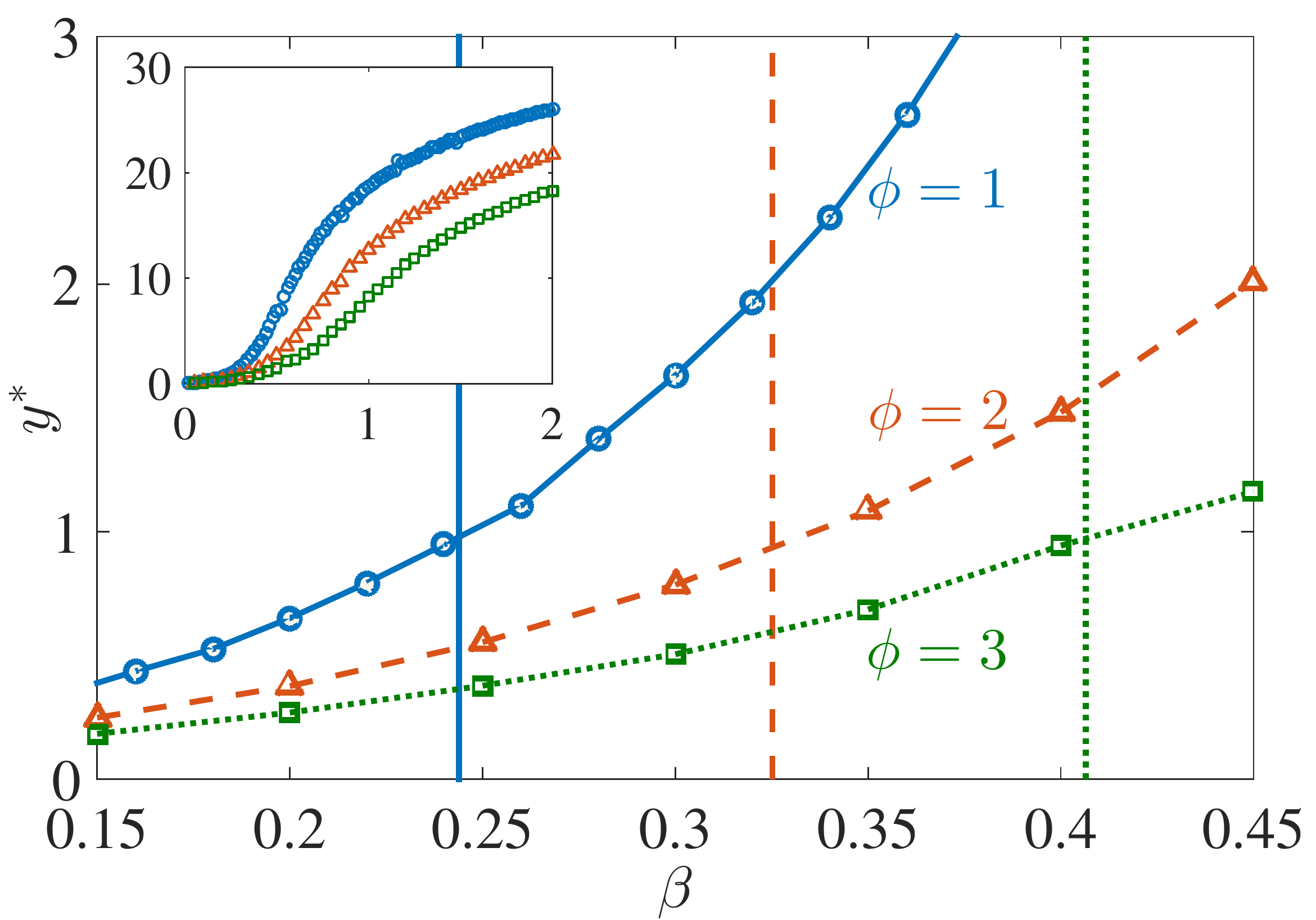}
\put(-163,100){1)}
\hspace{0.02\textwidth}
\includegraphics[width=.31\textwidth]{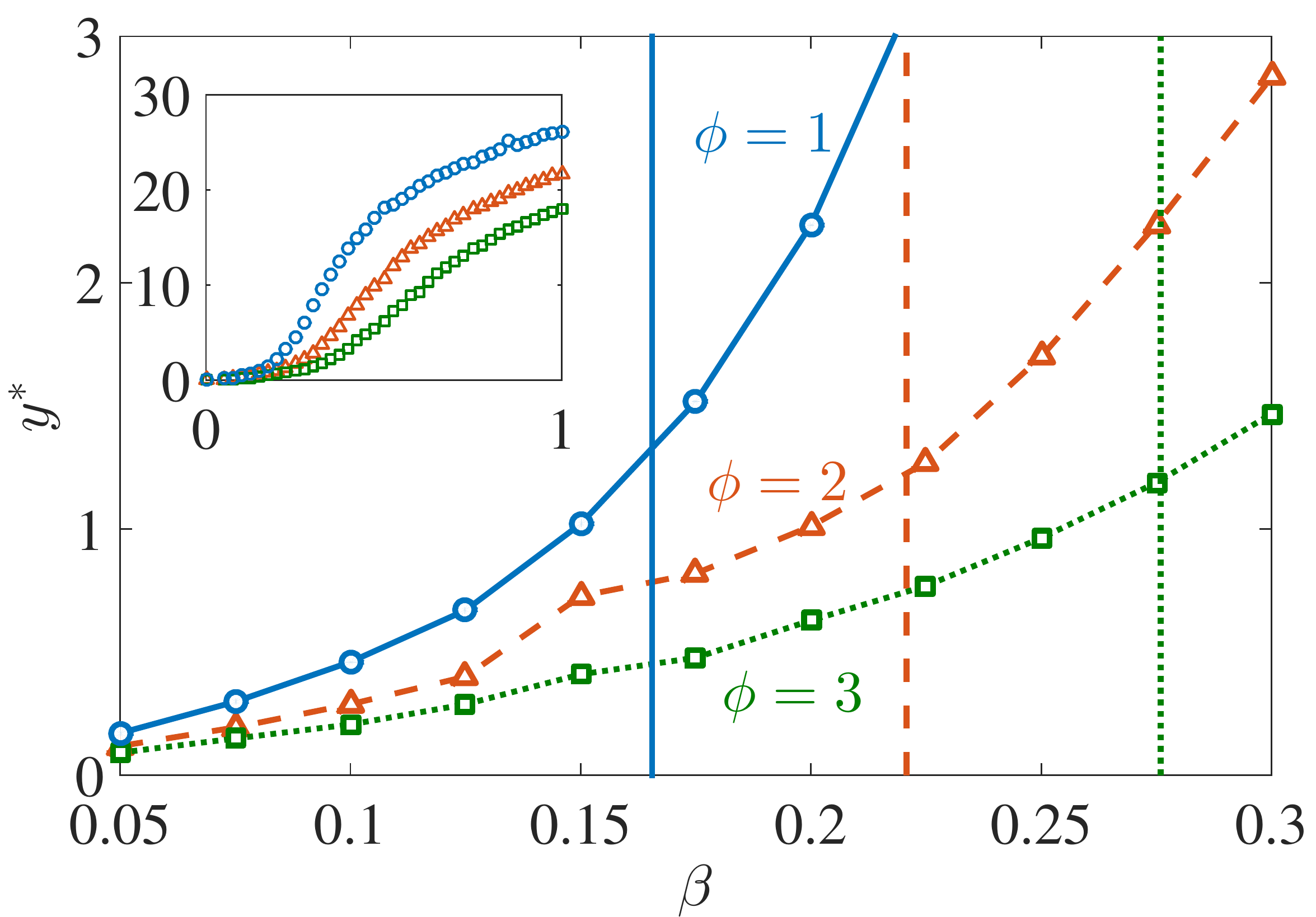}
\put(-160.5,100){2)}
\hspace{0.02\textwidth}
\includegraphics[width=.31\textwidth]{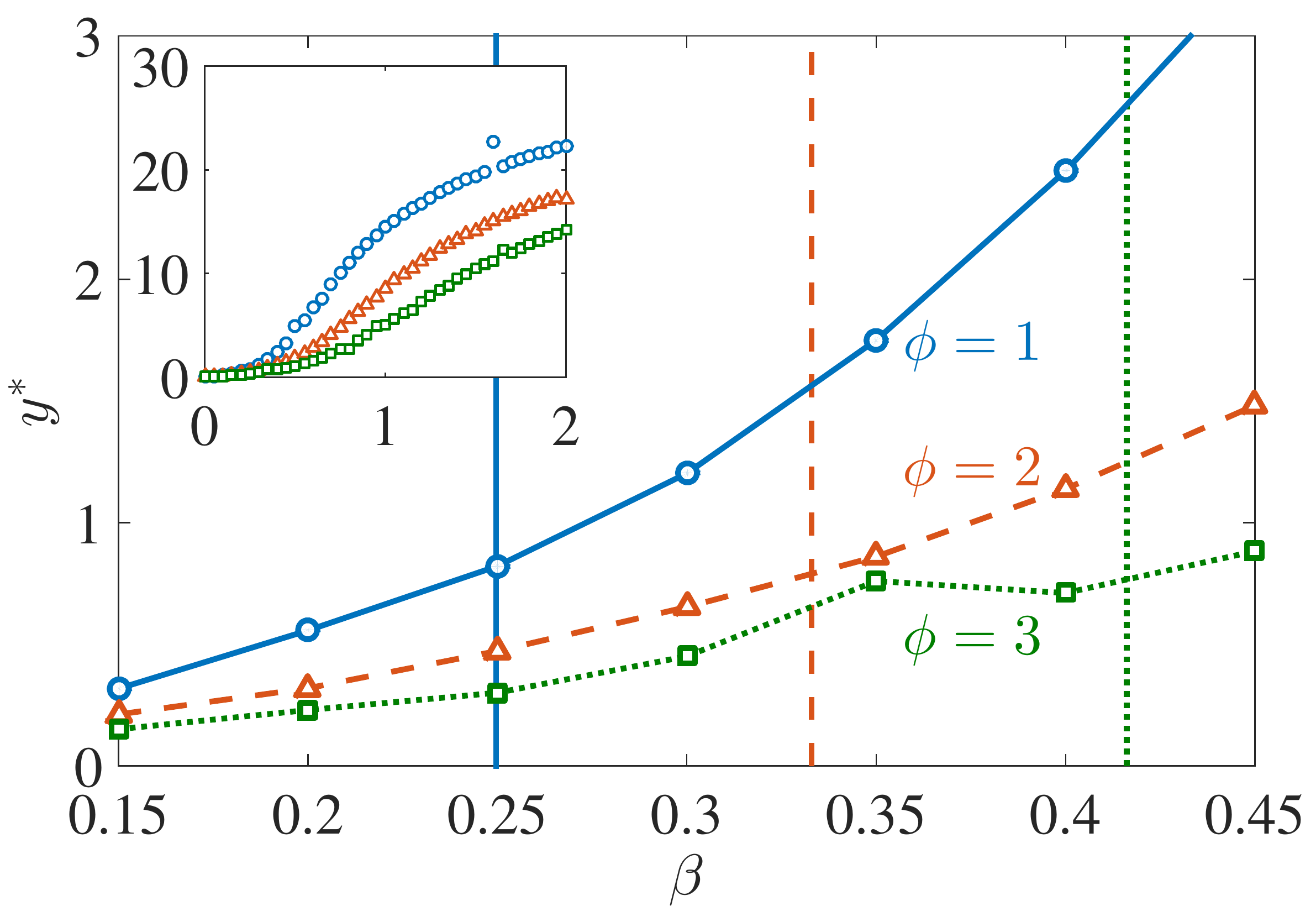}
\put(-160.5,100){3)}
\caption{Metastable number of infected nodes versus $\beta$ for $\phi = 1, 2, 3$ with $\delta = \psi = 1$. 1) Erd\H{o}s-R\'enyi graph ($p=0.1$), 2) Erd\H{o}s-R\'enyi graph ($p=0.2$), and 3) Barab\`asi-Albert model.}
\label{fig:meta_vs_beta}
\end{figure*}

Let the initial graph~$\mathcal G(0)$ be realizations of the following random
graph models with $n=40$ nodes: \emph{1}) an Erd\H{o}s-R\'enyi graph with edge
probability~$p = 0.1$, \emph{2}) an Erd\H{o}s-R\'enyi graph with edge
probability~$p = 0.2$, and \emph{3}) a Barab\`asi-Albert random graph with
average degree $3.65$. We fix the recovery rate to $\delta = 1$ for all nodes in
the graph, for the purpose of illustration. For three different values of the
reconnecting rate, $\psi \in \{1/2,1,2\}$, we show in Table~\ref{table:new} the
contour plots of the metastable number~$y^*$ of infected nodes as we vary the
values of the infection rate $\beta$ and the cutting rate $\phi$. We see how the
analytical \blue{lower bound}s (represented as dashed straight lines in the
figures in Table~\ref{table:new}) are in good accordance with the numerical
contour corresponding~$y^* = 1$ (orange \blue{thick} curves), in particular for
the  Erd\H{o}s-R\'enyi graph with edge probability~$p = 0.1$. For the specific
case of $\psi=1$, Figure~\ref{fig:meta_vs_beta} shows the values of $y^*$ when
$\phi = 1$ (\blue{blue line}), $2$ (\blue{orange dashed line}), and $3$
(\blue{green dotted line}), and $\beta$ varies from $0$ to $2$. The vertical
lines in the figure show the theoretical epidemic threshold values of~$\beta$
predicted from the \blue{lower bound}~\eqref{eq:threshold:homo}. We confirm that
these thresholds are in good accordance with the numerical thresholds, which
corresponds to the values for which the curves cross the horizontal line~$y^* =
1$.

\section{Cost-Optimal Adaptation for Epidemic Eradication} 

Based on our theoretical results in the last section, in this section we study
the problem of tuning the rates of the heterogeneous ASIS model in order to
eradicate an epidemic outbreak. Specifically, we consider the situation where we
can tune the values of the infection, recovery, and cutting rates in the network
(for technical reasons, we cannot tune the reconnecting rates in our framework,
as we discuss at the end of Appendix~\ref{appx:gp}). In this setup, we assume
that there is a cost associated with tuning the values of these rates. These
costs are described using the following collection of cost functions. The first
cost function $f(\beta)$ accounts for the cost of tuning the infection rates in
the network to the values in the vector $\beta = (\beta_i)_i$. In other words,
if we want to have a network where the infection rates are those in the vector
$\beta$, we need to pay $f(\beta)$ monetary units. Similarly, the functions
$g(\delta)$ and $h(\phi)$ account for the cost of tuning the recovery and the
cutting rates to the vectors $\delta = (\delta_i)_i$ and $\phi = (\phi_{ij})_{i,
j}$, respectively. Using these cost functions, our objective is to find the
cost-optimal investment profile for tuning these rates in order to eradicate the
disease at a desired exponential decay rate. From our theoretical analysis, this
exponential decay rate is given by $\lambda_{\max}(M)$ in \eqref{eq:thr}. Hence,
the optimal tuning problem can be stated as follows:
\begin{problem}[Cost-optimal eradication]\label{prb:}
Given a desired exponential decay rate $\bar \lambda > 0$, and positive numbers $\ubar
\beta$, $\bar \beta$, $\ubar{\delta}$, $\bar{\delta}$, $\ubar \phi$, and $\bar
\phi$, find the set of rates $(\beta_i)_i$, $(\delta_i)_i$, and $(\phi_{ij})_{i,j}$ satisfying the following feasibility bounds
\begin{gather}\label{eq:bounds1}
\ubar{\beta}\leq \beta_{i} \leq \bar{\beta},\:
\ubar{\delta}\leq \delta_{i} \leq \bar{\delta}, \:
\ubar{\phi}\leq \phi_{ij} \leq \bar{\phi}, \:
\end{gather}
such that the infection probabilities $p_i$ in the heterogeneous ASIS model decay to zero exponentially fast at a rate~$\bar  \lambda$, while the total tuning cost
\begin{equation*}
C = f(\beta) + g(\delta) + h(\psi)
\end{equation*}
is minimized.
\end{problem}

In what follows, we show how this problem can be cast into a type of
optimization problems called geometric programs~\cite{Boyd2007}, which allows us
to find the cost-optimal rates in polynomial time. The techniques herein
presented extend those in~\cite{Preciado2013,Preciado2014}, where the authors
proposed the use of convex programming to find the cost-optimal allocation of
resources to eradicate an epidemic outbreak in arbitrary \emph{static} networks.
The techniques presented below work for a wide family of cost functions, called
posynomial functions  (see \cite{Boyd2007} for more details). For simplicity in
our exposition, we illustrate the idea behind our approach with these particular
cost functions:
\begin{equation*}
\begin{aligned}
f(\beta) &= c_1 + c_2 \sum_{i=1}^n \frac{1}{\beta_i^{p_i}}, 
  \\
g(\delta) &= c_3 + c_4 \sum_{i=1}^n \frac{1}{(q_i-\delta_i)^{r_i}}, 
\\
h(\phi) &= c_5 + c_6 \sum_{\{i, j \}\in\mathcal G(0)}\frac{1}{(s_{ij} - \phi_{ij})^{u_{ij}}}, 
\end{aligned}
\end{equation*}
where $c_1$,$\dotsc$, $c_6$ are a set of parameters that are chosen to normalize the cost functions to satisfy the following equalities:
\begin{equation*}
\begin{aligned}
f(\bar \beta, \dotsc, \bar \beta) & =0, 
& f(\ubar \beta, \dotsc, \ubar \beta) & =1, 
\\
g(\bar \delta, \dotsc, \bar \delta) & =1, 
& g(\ubar \delta, \dotsc, \ubar \delta) &=0, 
\\
h(\bar \phi, \dotsc, \bar \phi) &=1, 
& h(\ubar \phi, \dotsc, \ubar \phi) & =0,
\end{aligned}
\end{equation*}
and the constants $p_i$, $q_i$, $r_{i}$, $s_{ij}$, and $u_{ij}$ are positive
real parameters that can be used to modify the shape of the cost functions. The
parameters in $f$, the function representing the cost of tuning the infection
rates, are chosen to make the function decreasing with respect to each
$\beta_i$. In other words, the higher the tuning investment, the smaller the
resulting infection rate (as we should expect in practical situations).
Following a similar reasoning, the functions~$g$ and~$\phi$ are set to be
increasing.

\begin{table*}[tb]
\newcommand{\figwidth}{.25\textwidth}
\newcommand{\gwidth}{.11\textwidth}
\newcommand{\gheight}{4.1cm}
\caption{Cost-optimal cutting rates $\phi_{ij}$}
\label{table:}
\begin{tabular}{cccc}
\hline	
\parbox[c][2.5em]{\gwidth}{Horizontal\\axis}& \parbox[c][2.5em]{\figwidth}{a) Product of degrees\\of $i$ and $j$} & \parbox[c]{\figwidth}{b) Product of eigenvector\\centralities of $i$ and $j$} & c) Edge betweenness of $\{i, j\}$
\\
\hhline{====}
\parbox[c][\gheight]{\gwidth}{1) Erd\H{o}s-\\R\'enyi\\graph} &
\parbox[c]{\figwidth}{\includegraphics[width=\figwidth]{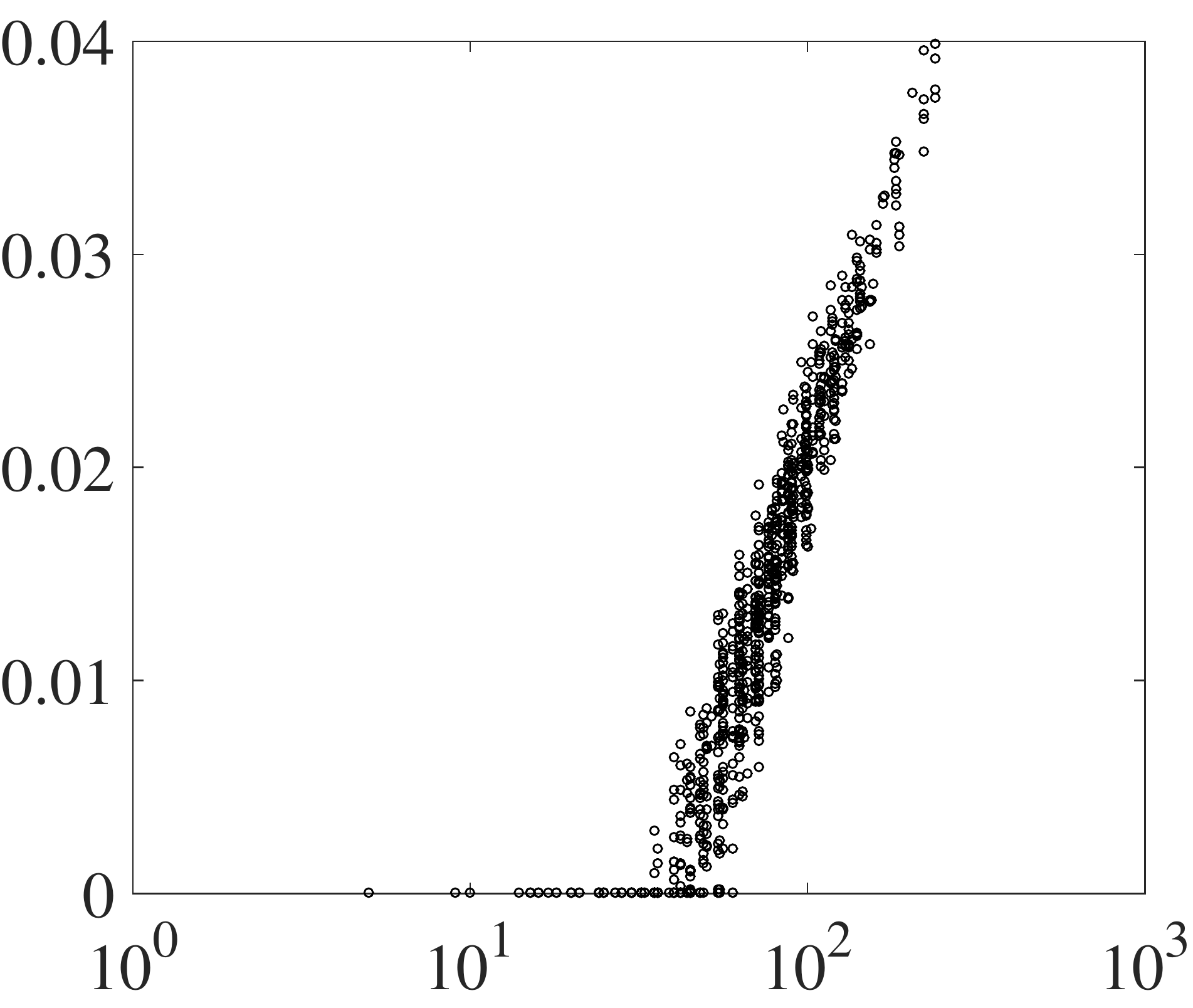}}&
\parbox[c]{\figwidth}{\includegraphics[width=\figwidth]{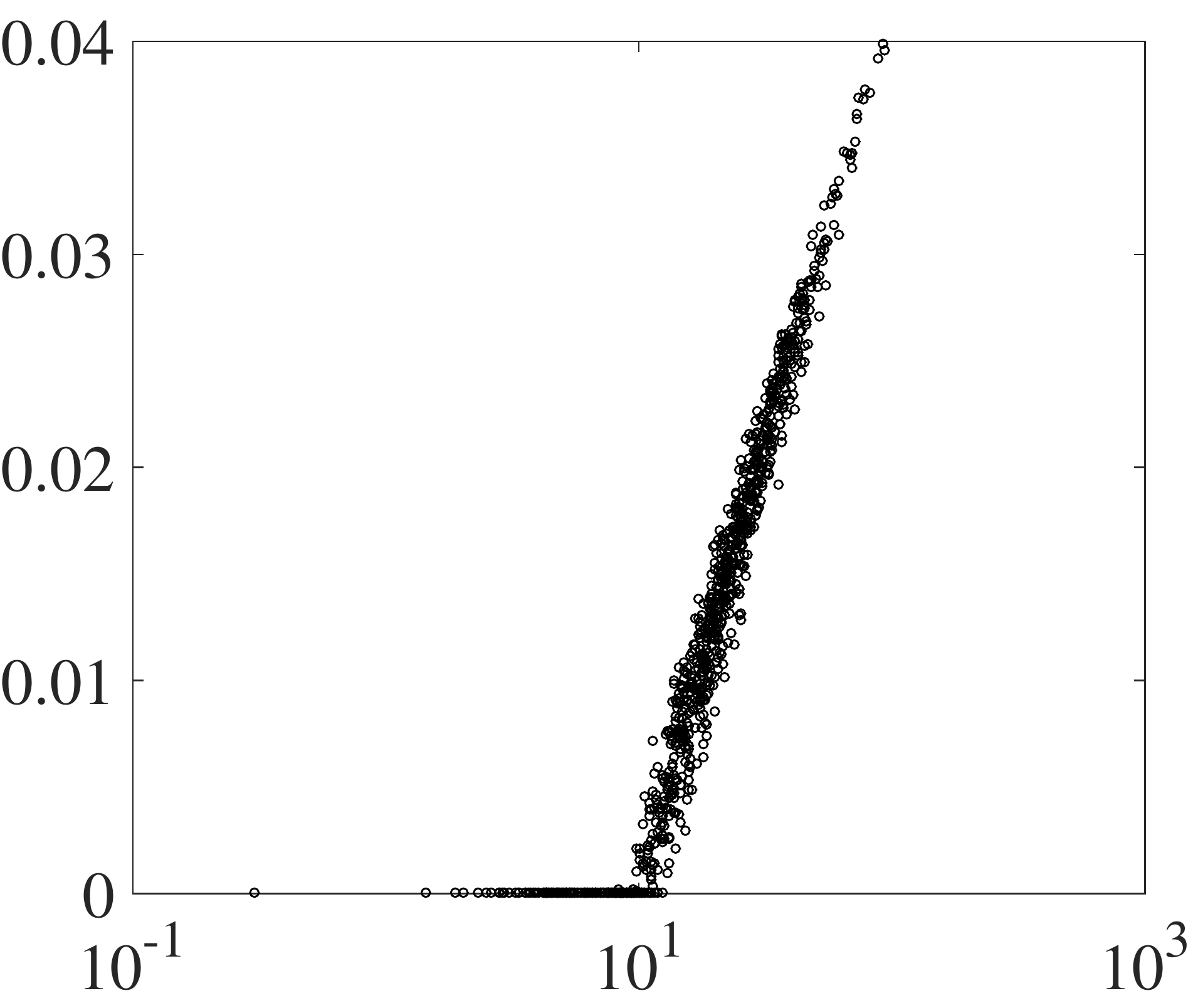}} &
\parbox[c]{\figwidth}{\includegraphics[width=\figwidth]{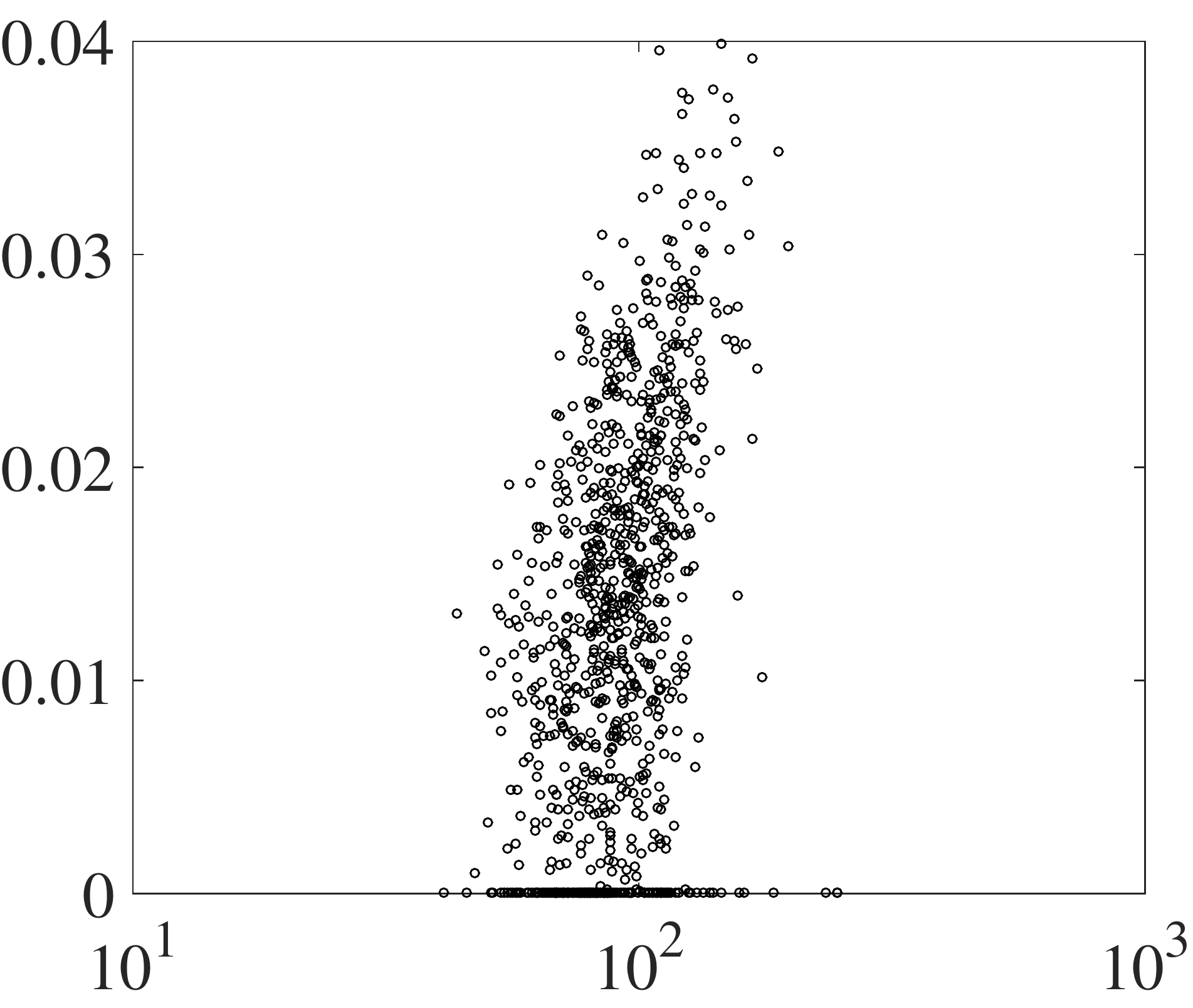}}
\\
\hline
\parbox[c][\gheight]{\gwidth}{2) Barab\`asi-\\Albert\\model} &
\parbox[c]{\figwidth}{\includegraphics[width=\figwidth]{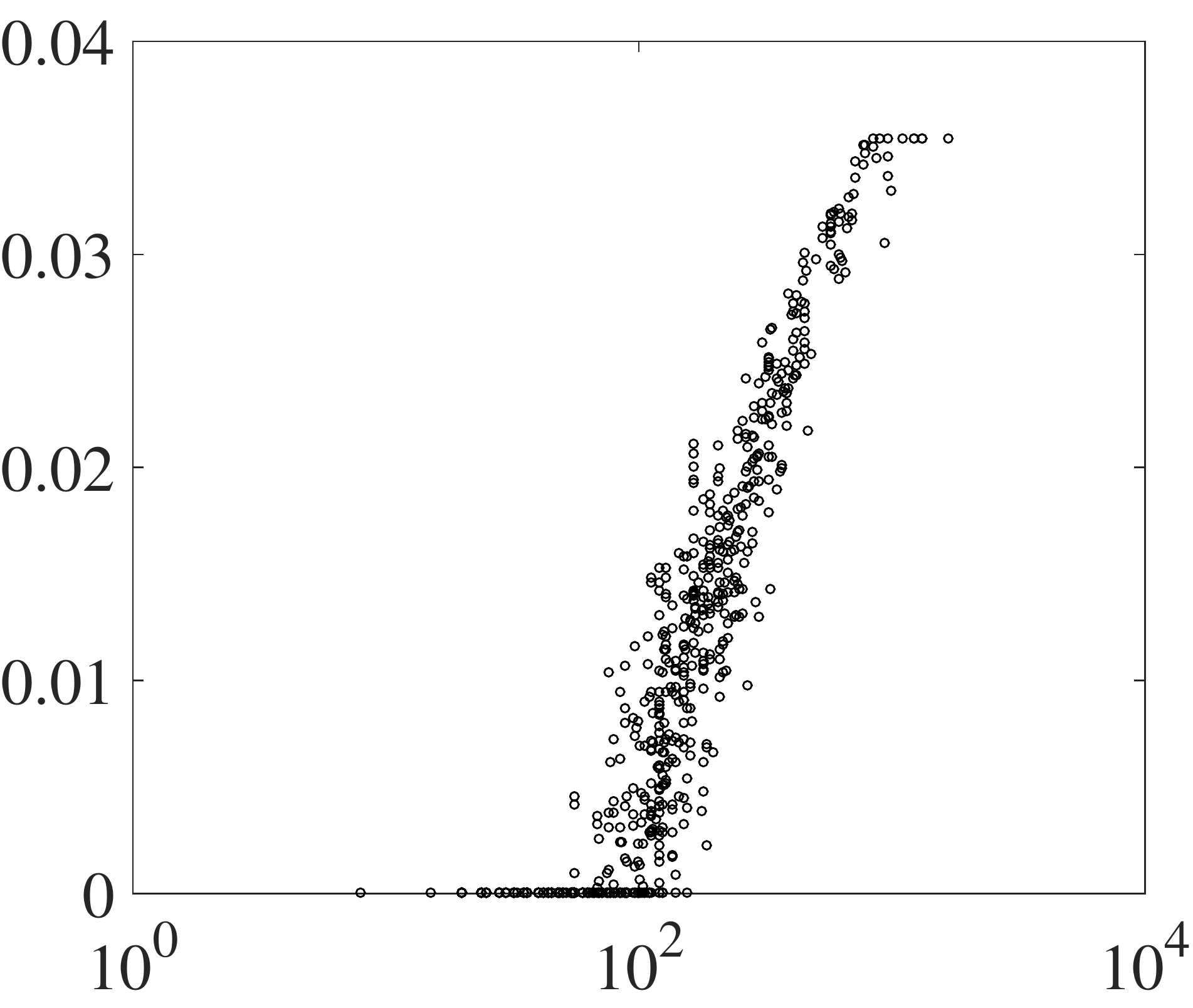}}&
\parbox[c]{\figwidth}{\includegraphics[width=\figwidth]{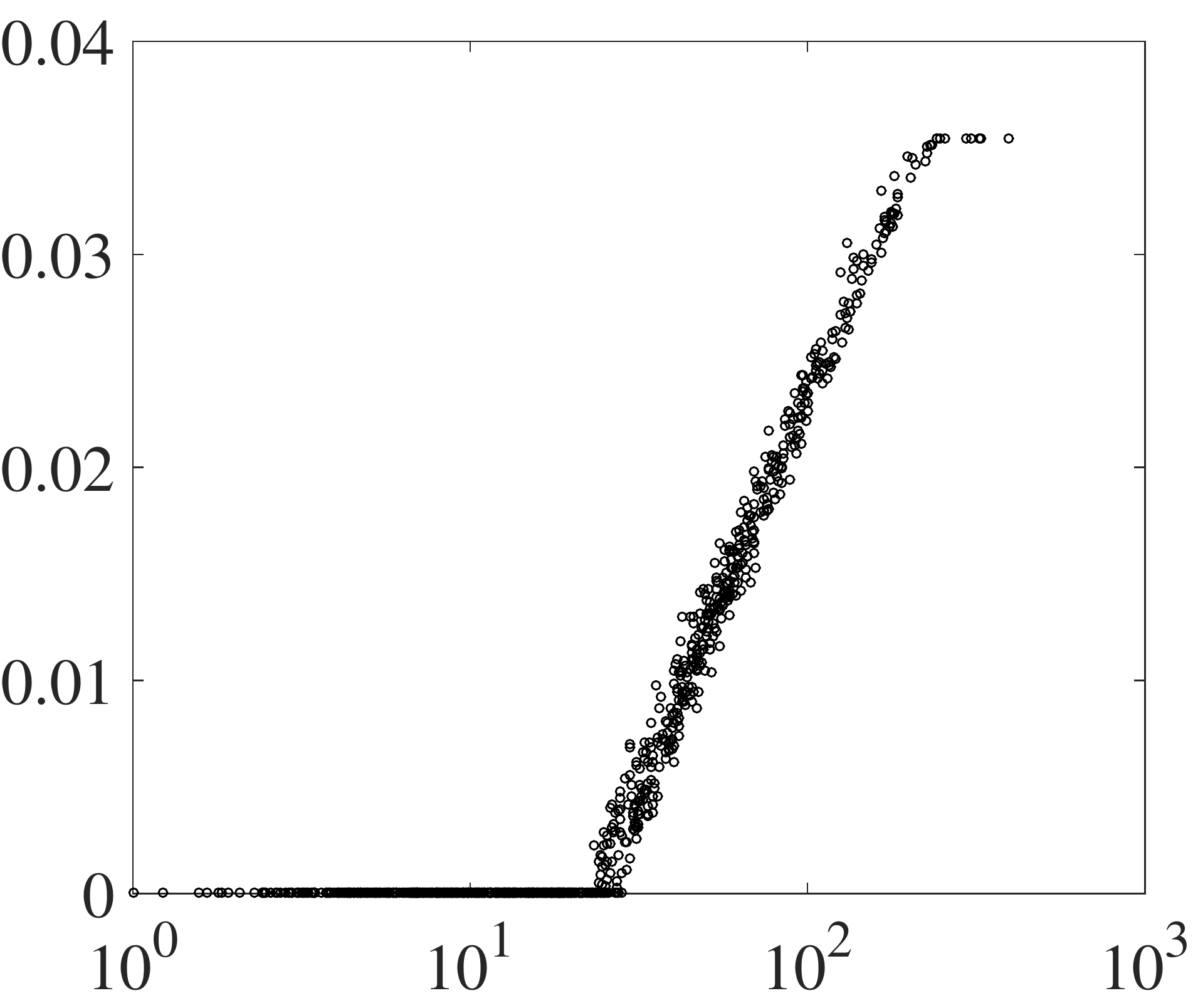}} &
\parbox[c]{\figwidth}{\includegraphics[width=\figwidth]{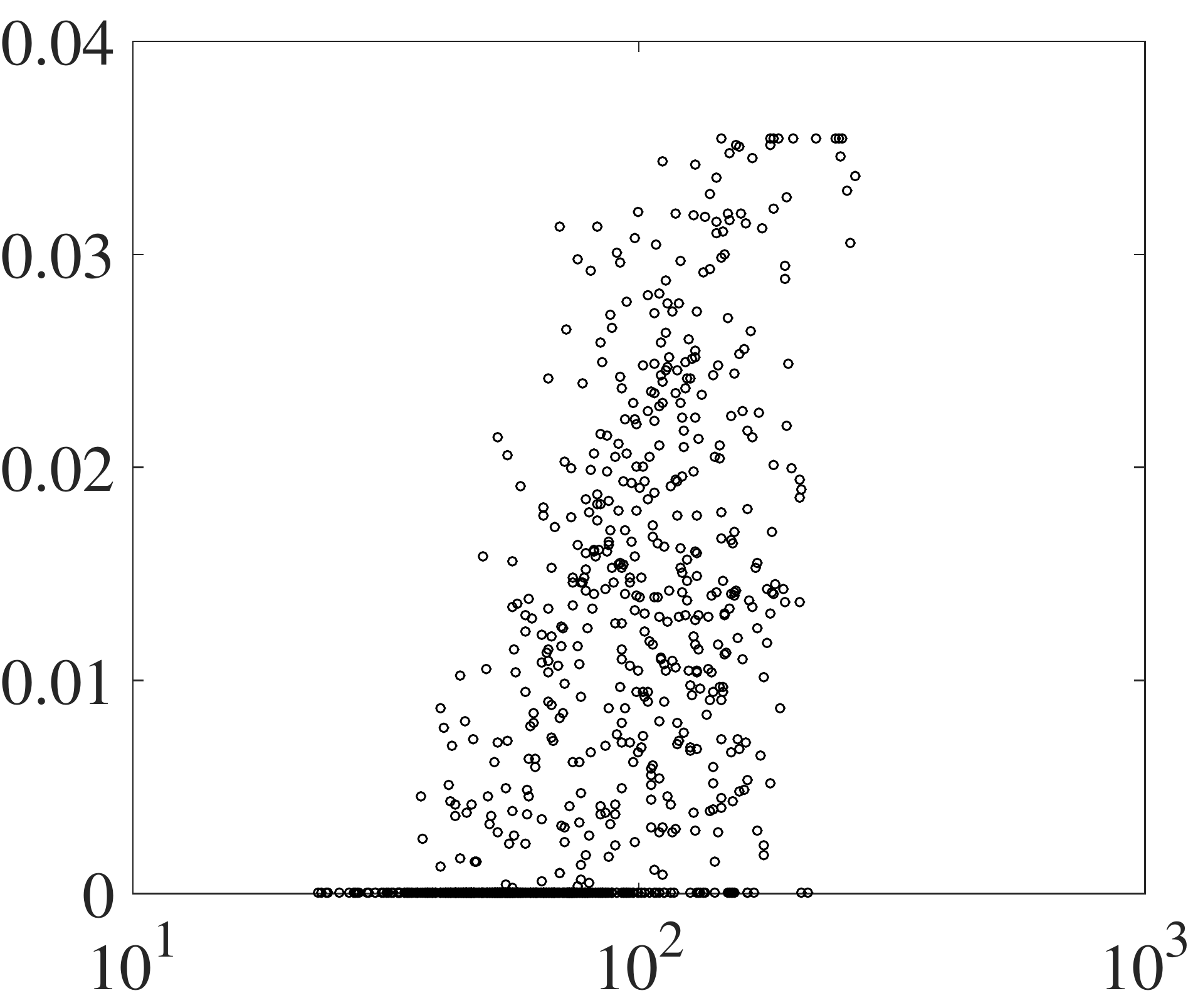}}
\\
\hline
\parbox[c][\gheight]{\gwidth}{3) Facebook\\network} &
\parbox[c]{\figwidth}{\includegraphics[width=\figwidth]{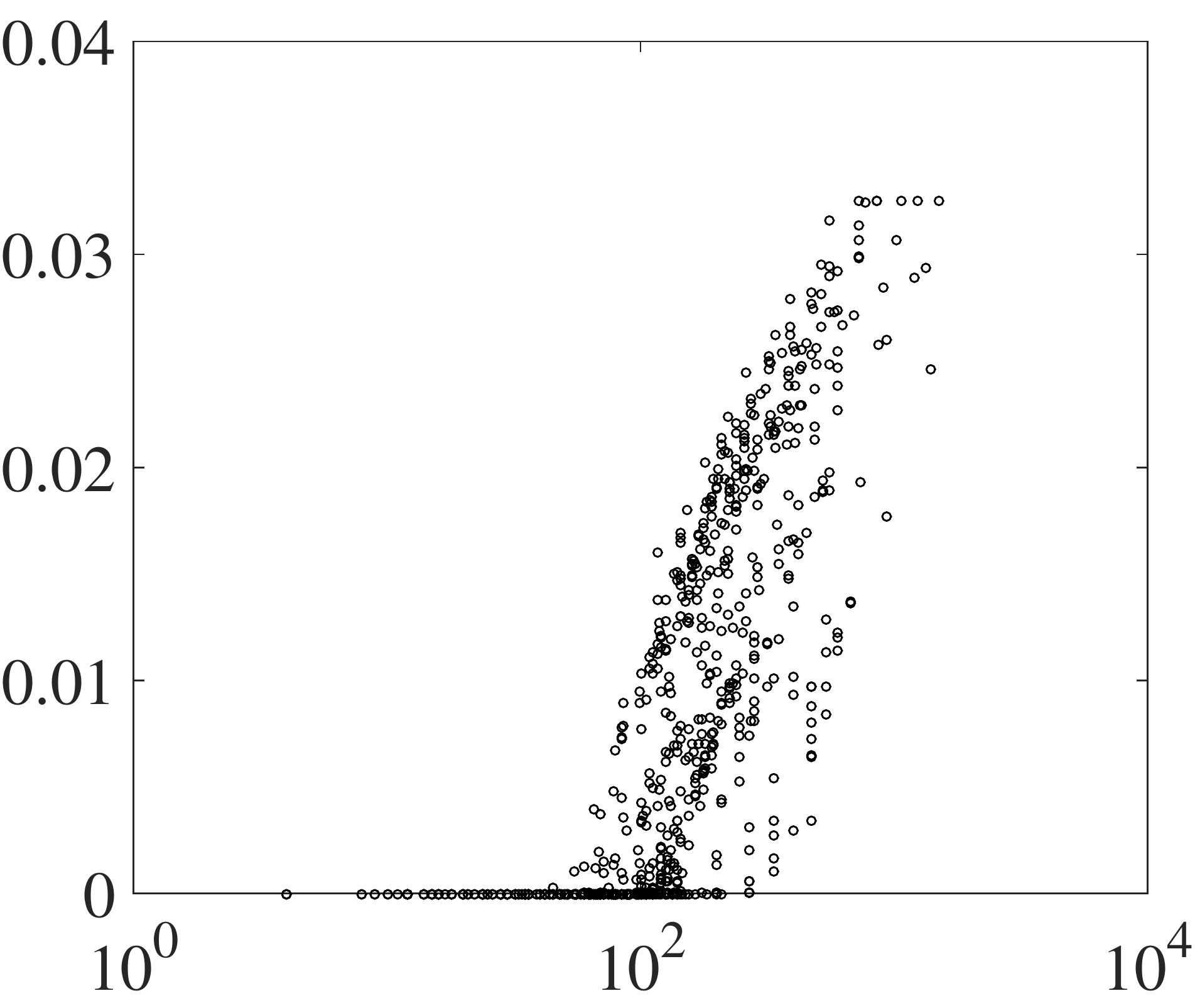}}&
\parbox[c]{\figwidth}{\includegraphics[width=\figwidth]{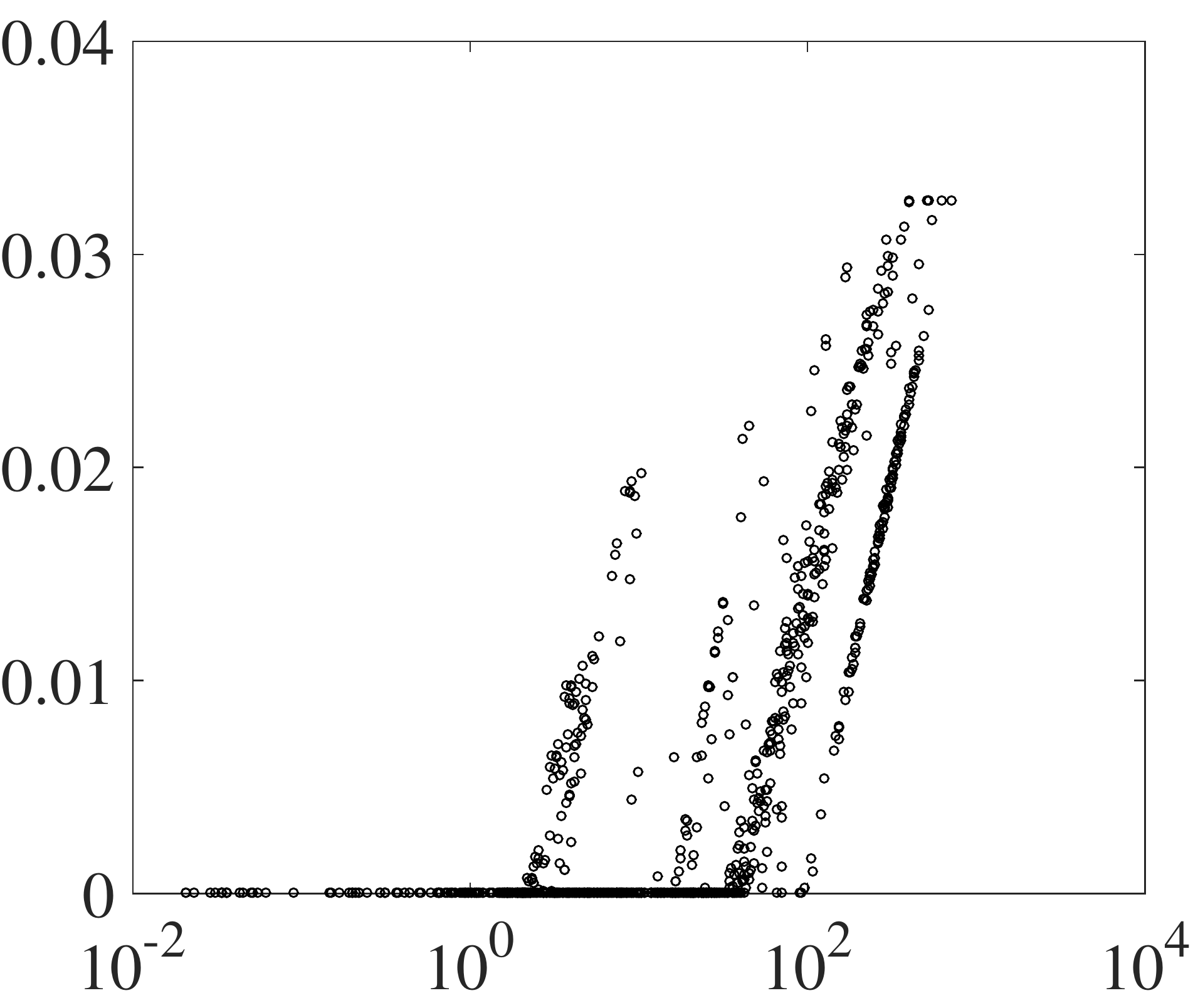}} &
\parbox[c]{\figwidth}{\includegraphics[width=\figwidth]{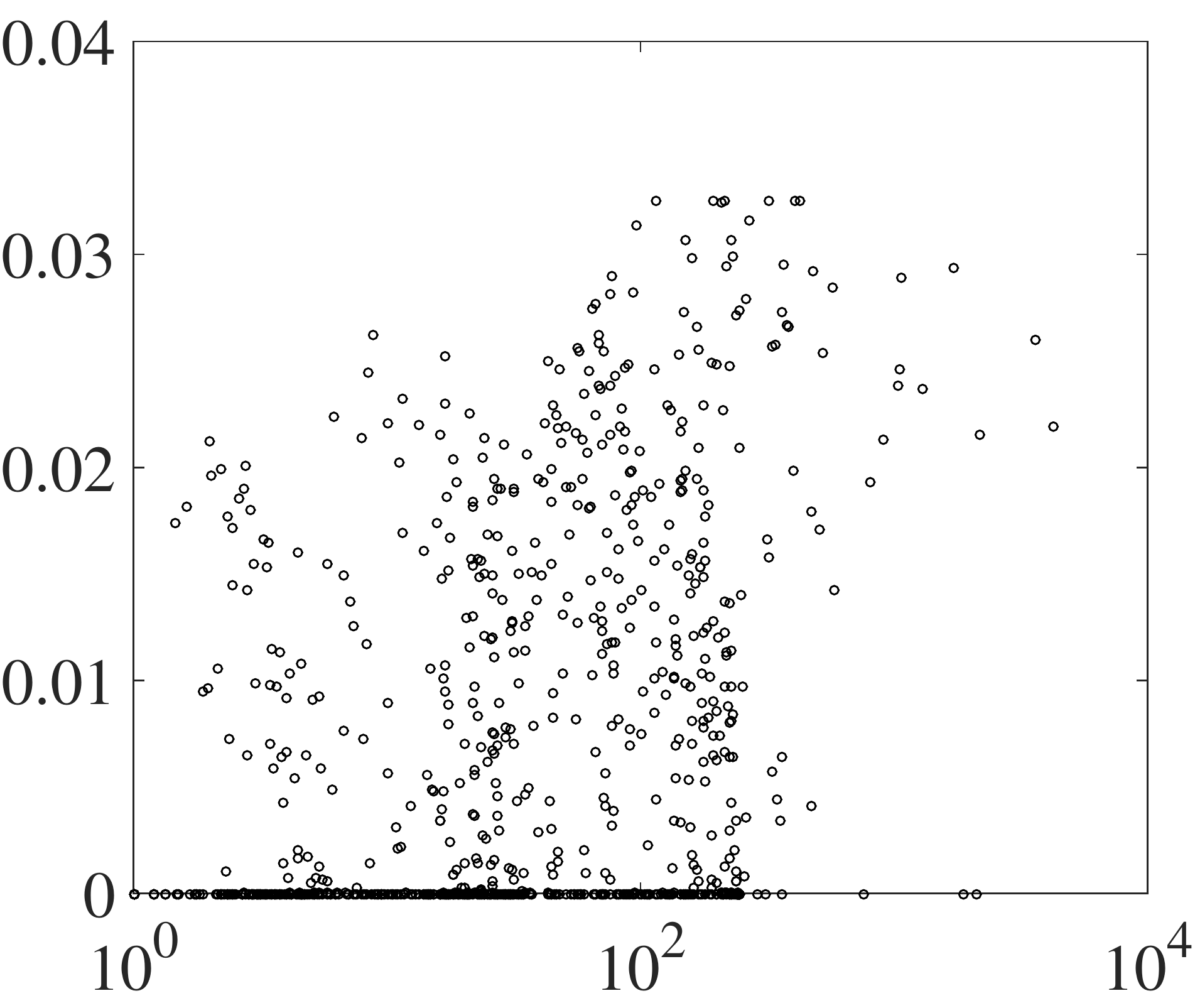}}
\\
\hline
\end{tabular}
\end{table*}

Once the cost functions are selected, we must solve the problem of finding the optimal tuning investment to achieve a desired exponential decay rate in the probabilities of infection. From the inequality in \eqref{eq:d[pq]/dt}, the infection probabilities $p_1$, $\dotsc$, $p_n$  decay exponentially at a rate  (at least) $\bar \lambda$ if 
\begin{equation}\label{eq:decay_condition}
\lambda_{\max}(M) \leq -\bar \lambda. 
\end{equation}
Since $M$ is an irreducible and Metzler matrix, we can use Perron-Frobenius
theory~\cite{Horn1990} to prove that \eqref{eq:decay_condition} is satisfied if
there exists an entry-wise positive vector~$v$ satisfying the following
entry-wise vector inequality (see \cite{Preciado2014} for more details):
\begin{equation}\label{eq:Mv<-lambdav}
Mv < -\bar \lambda v.
\end{equation}
Therefore, Problem~\ref{prb:} can be reduced to the
following equivalent optimization problem:
\begin{equation}\label{eq:optim}
\begin{aligned}
\minimize_{\beta,\,\delta,\,\phi,\,v}\ \ 
&f(\beta) + g(\delta) + h(\phi) 
\\
\subjectto\ \,
&\text{\eqref{eq:bounds1}, \eqref{eq:Mv<-lambdav}, and $v>0$.}
\end{aligned}
\end{equation}
As we show in Appendix~\ref{appx:gp}, we can \blue{equivalently} transform this optimization problem into a geometric program~\cite{Boyd2007}, which can be efficiently solved using standard optimization software. \blue{The computational complexity of solving the resulting geometric program is $O((n+m)^{7/2})$, where $n$ is the number of nodes and $m$ is the number of edges in the initial network $\mathcal G(0)$.}

In the rest of this section, we compute the optimal tuning profiles for three
different graphs and compare our results with several network centralities. In
our simulations, we consider the following three graphs with $n=247$ nodes:
\emph{1}) an Erd\H{o}s-R\'enyi graph with $916$ edges, \emph{2}) a
Barab\`asi-Albert random graph with $966$ edges, and \emph{3}) a social subgraph
(obtained from Facebook) with $947$ edges.  For simplicity in our simulations,
we assume that all nodes share the same recovery rate~$\delta = 0.1$ and
infection rate $\beta = \delta/(1.1 \rho)$, where $\rho$ denotes the spectral
radius of each initial graph. Since $\delta/\beta = 1.1\, \rho > \rho$, the
extinction condition \eqref{eq:threshold:homo} indicates that the infection
process does not necessarily die out without adaptation, i.e., when $\phi_{ij} =
0$. The rest of parameters in our simulations are set as follows: we let $\ubar
\phi = 0$, $\bar \phi = 4\beta$, and $\psi_{ij} = \beta$. Also, the parameters
in the cost functions are $p_i = q_i = r_i = u_{ij} = 1$ and $s_{ij} = 2\bar
\phi$. The desired exponential decay rate is chosen to be $\bar \lambda =
0.005$.

Using this set of parameters, we solve the optimization problem~\eqref{eq:optim}
following the procedure described in Appendix~\ref{appx:gp}. Our numerical
results are illustrated in various figures included in Table~\ref{table:}. Each
figure is a scatter plot where each point corresponds to a particular edge
$\{i,j\}\in \mathcal E(0)$; the ordinate of each point corresponding to its
optimal cutting rate $\phi_{ij}$, and the abscissa of each point corresponds to
a particular edge-centrality measure. In these figures, we use three different
edge-centrality measures: a) the product of the degrees of nodes $i$ and $j$
(left column), b) the product of the eigenvector-centralities of nodes $i$ and
$j$ (center column), and c) the betweenness centralities of edge $\{i, j\}$
(right column).

In our numerical results, we observe how both degree-based and eigenvector-based
edge-centralities are good measures for determining the amount of investment in
tuning cutting rates. In contrast, betweenness centrality does not show a
significant dependency on the optimal cutting rates. In particular, for the
synthetic networks in rows 1) and 2) in Table \ref{table:}, we observe an almost
piecewise affine relationship with the centrality measures in columns a) and b).
In particular, in these subplots we observe how edges of low centrality require
no investment, while for higher-centrality edges, the tuning investment tends to
increase affinely as the centrality of the edge increases -- as expected. For
the real social network in row c), the relationship between investment and
centralities is still strong -- although not as clear as in synthetic networks.
In the scatter plot corresponding to the relationship between the optimal
investment and the eigenvector-based centrality in the real social network 
(lower center figure in Table  \ref{table:}), we observe a collection of several
stratified parallel lines. We conjecture that each line corresponds to a
different community inside the social network; in other words, the relationship
between centrality and optimal investment is almost affine inside each
community.

\section{Conclusion}

We have studied an \emph{adaptive} susceptible-infected-susceptible (ASIS) model
with heterogeneous node and edge dynamics and arbitrary network topologies. We
have derived an explicit expression for a \blue{lower bound} on the epidemic
threshold of this model in terms of the maximum real eigenvalue of a matrix that
depends explicitly on the network topology and the parameters of the model. For
networks with homogeneous node and edge dynamics, the \blue{lower bound} turns
out to be a constant multiple of the epidemic threshold in the standard SIS
model over static networks (in particular, the inverse of the spectral radius).
Furthermore, based on our results, we have proposed an optimization framework to
find the cost-optimal adaptation rates in order to eradicate the epidemics. We
have confirmed the accuracy of our theoretical results with several numerical
simulations and compare cost-optimal adaptation rates with popular centrality
measures in various networks.

%

\appendix

\section{Irreducibility of $M$}\label{appx:pf:irreducibility}

We show that the matrix $M$ defined in \eqref{eq:defM} is irreducible, that is, 
there is no similarity transformation that transforms $M$ into a block
upper-triangular matrix. For this purpose, define
\begin{equation*}
L = \begin{bmatrix}
O & T \\ J&S
\end{bmatrix}, 
\end{equation*}
where $J = \bigoplus_{i=1}^n\onev_{d_i}$, $T = \col_{1\leq i\leq n}T_i$, and $S
= \col_{1\leq i\leq n}(\onev_{d_i}\otimes T_i)$. Since the rates $\beta_i$ and
$\psi_{ij}$ are positive, if $M_{ij} = 0$, then $L_{ij} = 0$ for all distinct
$i$ and $j$. Therefore, to prove the irreducibility of $M$, it is sufficient to
show that $L$ is irreducible.

In order to show that $L$ is irreducible, we shall show that the directed graph
$\mathcal H$ on the nodes $1, \dotsc, n+2m$, defined as the graph having
adjacency matrix~$L\in \mathbb{R}^{(n+2m)\times (n+2m)}$, is strongly connected.
We identify the nodes $1$, $\dotsc$, $n+2m$ and variables $p_1$, $\dotsc$,
$p_n$, $q_{1j}$ ($j\in\mathcal{N}_1(0)$), $\dotsc$, $q_{nj}$
({$j\in\mathcal{N}_n(0)$}). Then, the upper-right block~$T$ of the matrix $L$
indicates that the graph $\mathcal H$ contains the directed edge~$(p_i, q_{ji})$
for all $i=1, \dotsc, n$ and $j \in \mathcal N_i(0)$. Similarly, from the
matrices $J$ and~$S$ in $M$, we see that $\mathcal H$ contains the directed
edges~$(q_{ij}, p_i)$ and $(q_{ij}, q_{ki})$ for all $i=1, \dotsc, n$ and {$j, k
\in \mathcal N_i(0)$}.

Using these observations, let us first show that $\mathcal H$ has a directed
path from $p_i$ to $p_j$ for all $i, j\in \{1, \dotsc, n\}$. Since $\mathcal
G(0)$ is strongly connected, it has a path $(i_0, \dotsc, i_\ell)$ such that
$i_0 = i$ and $i_\ell = j$. Therefore, from the above observations, we see that
$\mathcal H$ contains the directed path $(p_i, q_{i_1, i_0}, q_{i_2, i_1},
\dotsc, q_{i_\ell, i_{\ell-1}}, p_j)$. In the same way, we can show that
$\mathcal H$ also contains the directed path $(p_i, q_{ji}, q_{ij}, p_i)$ for
every $\{i, j\} \in \mathcal E(0)$. These two types of directed paths in
$\mathcal H$ guarantees that $\mathcal H$ is strongly connected. Hence the
matrix~$L$ is irreducible, as desired.

\section{Derivation of \eqref{eq:threshold:homo}} \label{appx:pf:thm:stbl:homo}

In the homogeneous case, the matrix $M$ takes the form
\begin{equation*}
M = 
\begin{bmatrix}
-\delta I	& \beta T
\\
\psi J		& \beta S - (\delta + \phi + \psi) I
\end{bmatrix}.
\end{equation*}
In what follows, we show that $\lambda_{\max}(M)<0$ if and only if \eqref{eq:threshold:homo} holds true. 

Since $\mathcal G(0)$ is strongly connected by assumption, its adjacency
matrix~$A(0)$ is irreducible and therefore has an entry-wise positive
eigenvector~$v$ corresponding to the eigenvalue~$\rho$ (see~\cite{Farina2000}).
Define the positive vector $w = \col_{1\leq i\leq n}(v_i \onev_{d_i})$. Then,
the definition of~$T_i$ in \eqref{eq:def:Ti} shows $T_iw = \sum_{k\in\mathcal
N_i(0)} w_{ki} = \sum_{k\in\mathcal N_i(0)} v_k = (Av)_i = \rho v_i$ and
therefore $Tw = \lambda v$. In the same manner, we can show that {$Sw = \rho
w$}. Moreover, it is straightforward to check that $Jv = w$. Therefore, for a
real number $c$, it follows that
\begin{equation*} 
M
\begin{bmatrix}
cv\\w
\end{bmatrix}
=
\begin{bmatrix}
(\beta\rho - c\delta)v
\\
\left(c\psi + \beta \rho - (\delta + \phi + \psi)\right) w
\end{bmatrix}. 
\end{equation*}
Hence, if a real number $\lambda$ satisfies the following equations:
\begin{equation}\label{eq:uations}
\beta\rho - c\delta = c\lambda,\ 
c\psi + \beta \rho - (\delta + \phi + \psi) = \lambda, 
\end{equation}
then $\col(cv,w)$ is an eigenvector of $M$. Since $M$ is irreducible (shown in
Appendix~\ref{appx:pf:irreducibility}), by Perron-Frobenius
theory~\cite{Horn1990}, if $c>0$ then $\lambda_{\max}(M) = \lambda$ (see
\cite[Theorem~17]{Farina2000}). This, in particular, shows that
$\lambda_{\max}(M)<0$ if and only if there exist $c>0$ and $\lambda < 0$ such
that \eqref{eq:uations} holds.

The two equations  in \eqref{eq:uations} have two pairs of solutions $(c_1,
\lambda_1)$ and $(c_2, \lambda_2)$ such that $c_1<0, \lambda_1<0$, and $c_2>0$.
Therefore, we need to show $\lambda_2 < 0$ if and only if
\eqref{eq:threshold:homo} holds true. To see this, we notice that $\lambda_1$
and $\lambda_2$ are the solutions of the quadratic equation $\lambda^2 +
(2\delta+\phi+\psi-\beta\rho)\lambda + \delta(\delta+\phi+\psi) -
\beta\rho(\delta+\psi) = 0$ following from \eqref{eq:uations}. Since
$\lambda_1<0$, we have $\lambda_2 <0$ if and only if the constant term
$\delta(\delta+\phi+\psi) - \beta\rho(\delta+\psi)$ of the quadratic equation is
positive, which is indeed equivalent to \eqref{eq:threshold:homo}. This
completes the proof of the extinction condition stated
in~\eqref{eq:threshold:homo}.

\section{Geometric Programming}\label{appx:gp}

We first give a brief review of  geometric programs~\cite{Boyd2007}. Let $x_1$,
$\dotsc$, $x_n$ denote positive variables and define $x = (x_1, \dotsc, x_n)$.
We say that a real function~$g(x)$ is a {\it monomial} if there exist $c \geq 0$
and $a_1, \dotsc, a_n \in \mathbb{R}$ such that $g(x) = c x_{\mathstrut
1}^{a_{1}} \dotsm x_{\mathstrut n}^{a_n}$. Also, we say that a function~$f(x)$
is a {\it posynomial} if it is a sum of monomials of~$x$ (we point the readers
to~\cite{Boyd2007} for more details). Given a collection of posynomials
$f_0{(x)}$, $\dotsc$, $f_p{( x)}$ and monomials $g_1{(x)}$, $\dotsc$,
$g_q{(x)}$, the optimization problem
\begin{equation*} 
\begin{aligned}
\minimize_{ x}\ 
&
f_0({ x})
\\
\subjectto\ 
&
f_i({ x})\leq 1,\quad i=1, \dotsc, p, 
\\
&
g_j({ x}) = 1,\quad j=1, \dotsc, q, 
\end{aligned}
\end{equation*}
is called a {\it geometric program}. A constraint of the form $f(x)\leq 1$ with
$f(x)$ being a posynomial is called a posynomial constraint. Although geometric
programs are not convex, they can be efficiently converted into
\blue{equivalent} convex optimization problems~\cite{Boyd2007}.

In the following, we rewrite the optimization problem~\eqref{eq:optim} into a
geometric program using the new variables $\tilde \delta_i = q_i - \delta_i$ and
$\tilde \phi_{ij} = s_{ij} - \phi_{ij}$. By \eqref{eq:bounds1}, these variables
should satisfy the constraints
\begin{gather}
q_i- \bar{\delta}\leq \tilde \delta_{i} \leq q_i - \ubar{\delta},\label{eq:newbounds1}
\\
s_{ij} - \bar{\phi} \leq \tilde \phi_{ij} \leq s_{ij} - \ubar{\phi}.\label{eq:newbounds2}
\end{gather}
Also, using these variables, the cost function $C$ can be written as
$C(\beta, \tilde \delta, \tilde \psi) = f(\beta)  + \tilde g(\tilde \delta)
+\tilde h(\tilde \psi)$, where 
\begin{equation*}
\tilde g(\tilde \delta) = c_3 + c_4 \sum_{i=1}^n \frac{1}{\tilde{\delta}_i^{r_i}}, \quad 
\tilde h(\tilde \phi) = c_5 + c_6 \sum_{\{i, j \}\in\mathcal G(0)}\frac{1}{\tilde{\phi}_{ij}^{u_{ij}}}
\end{equation*}
are posynomials. In order to rewrite the constraint~\eqref{eq:Mv<-lambdav}, we first define the matrices
\begin{gather*}
\tilde D_1 = \bigoplus_{i=1}^n \tilde \delta_i,\quad\tilde D_2 = \bigoplus_{i=1}^n (\tilde \delta_i I_{d_i}), \\
\tilde \Phi =
\bigoplus_{i=1}^n \bigoplus_{j \in \mathcal N_i} \tilde \phi_{ij}, 
\quad 
\tilde
\Psi_2 = \bigoplus_{i=1}^n \bigoplus_{j\in\mathcal N_i(0)} \tilde \psi_{ij}.
\end{gather*}
We also introduce the positive constants $\bar q = \max_{i}q_i$, $\bar{\psi} = \max_{i, j}\psi_{ij}$, $\bar s = \max_{ij}\bar s_{ij}$. Now, adding $(\bar q + \bar \psi + \bar s) v$ to both sides of \eqref{eq:Mv<-lambdav}, we equivalently obtain
\begin{equation}\label{eq:Mtildev<-lambdav}
\tilde M v < (\bar q + \bar \psi + \bar s) v, 
\end{equation}
where $\tilde M = M + (\bar q + \bar \psi + \bar s)I$ is given by
\begin{widetext}
\begin{equation*}
\tilde M = \begin{bmatrix}
\tilde D_1 + (\bar{q}I-Q_1) + (\bar{\psi} + \bar{s}) I & B_1
\\
\Psi_1 & B_2 + \tilde D_2 + (\bar q I - Q_2) + \tilde \Phi + (\bar sI - S) + (\bar \psi I - \Psi_2) + \lambda I
\end{bmatrix}.
\end{equation*}
\end{widetext}

Summarizing, we have shown that the optimization problem~\eqref{eq:optim} is
equivalent to the following optimization problem with (entry-wise) positive
variables:
\begin{equation}\label{eq:gp}
\begin{aligned}
\minimize_{\beta,\,\tilde{\delta}_i,\,\tilde{\phi}_{ij},\,v}\ \ &
f(\beta)  + \tilde g(\tilde \delta)
+\tilde h(\tilde \psi)
\\
\subjectto\,\,\,&
\text{\eqref{eq:bounds1}, \eqref{eq:newbounds1}, \eqref{eq:newbounds2}, and \eqref{eq:Mtildev<-lambdav}}.
\end{aligned}
\end{equation}
In this optimization problem, the objective function is a posynomial in the
variables $\beta$, $\tilde \delta$, and $\tilde \phi$. Also, the box constraints
\eqref{eq:bounds1}, \eqref{eq:newbounds1}, and \eqref{eq:newbounds2} can be
written as posynomial constraints~\cite{Boyd2007}. Finally, since each entry of
the matrix $\tilde M$ is a posynomial in the variables $\beta$, $\tilde \delta$,
and~$\tilde \phi$, the vector-constraint \eqref{eq:Mtildev<-lambdav} yields
$n+2m$ posynomial constraints. Therefore, the optimization problem \eqref{eq:gp}
is a geometric program, as desired. \blue{Furthremore, a standard estimate on
the computational complexity of solving geometric program (see, e.g.,
\cite[Proposition~3]{Ogura2015a}) shows that the computational complexity of
solving the optimization problem in \eqref{eq:gp} is given by $O((n+m)^{7/2})$.}

Finally we remark that $\tilde M$ contains both the terms~$\psi_{ij}$ and
$-\psi_{ij}$ so that we cannot use $\psi_{ij}$ as the decision variable in the
geometric program~\eqref{eq:gp} due to the positivity constraint on decision
variables. By this reason, we cannot design the reconnecting rates $\psi_{ij}$
under the framework presented in this paper. This issue is left as an open
problem.

\end{document}